\documentclass{aa}
\usepackage{times,epsfig,amssymb,amsmath}
\usepackage{natbib}
\bibpunct{(}{)}{;}{a}{}{,} 

\newcommand{\chandra}{{\it Chandra}}
\newcommand{\xmm}{{\it XMM-Newton}}
\newcommand{\asca}{{\it ASCA}}
\newcommand{\rosat}{{\it ROSAT}}

\newcommand\solar{\hbox{{$Z_{\odot}$}}}

\title{Apparent high metallicity in 3-4 keV galaxy clusters:\\
the inverse iron-bias in action in the case of the merging cluster Abell 2028}
\titlerunning{Inverse iron-bias in Abell~2028}

\author{F. Gastaldello\inst{1,2,3} \and S. Ettori\inst{4,5} \and I. Balestra\inst{6} \and F. Brighenti\inst{7,8} \and D.~A. Buote\inst{2} \and S. De Grandi\inst{9} \and S. Ghizzardi\inst{1} \and M. Gitti\inst{4,10} \and P. Tozzi\inst{11} 
} 
\authorrunning{F. Gastaldello et al.}

\institute{
INAF, IASF, via Bassini 15, I-20133 Milano, Italy
\and Department of Physics and Astronomy, University of California at Irvine, 4129, Frederick Reines Hall, Irvine, CA,92697-4575
\and Occhialini Fellow
\and INAF, Osservatorio Astronomico di Bologna, via Ranzani 1, I-40127 Bologna, Italy
\and INFN, Sezione di Bologna, viale Berti Pichat 6/2, I-40127 Bologna, Italy
\and MPE, Giessenbachstrasse, Postfach 1312, D-85741 Garching, Germany
\and University of California Observatories/Lick Observatory, University of California, Santa Cruz, CA 95064
\and Dipartimento di Astronomia, Universit\`a di Bologna, via Ranzani 1, Bologna 40127, Italy
\and INAF, Osservatorio Astronomico di Brera, via E. Bianchi 46, I-23807 Merate(LC), Italy
\and Harvard-Smithsonian Center for Astrophysics, 60 Garden Street, Cambridge, MA 02138, USA
\and INAF, Osservatorio Astronomico di Trieste, via G.B. Tiepolo 11, I-34131 Trieste, Italy
}

\offprints{F. Gastaldello}

\mail{gasta@lambrate.inaf.it}

\begin{document}

\abstract
{ Recent work based on a global measurement of the ICM properties finds evidence for an increase
in the iron abundance in galaxy clusters with temperatures around 2-4 keV up to a value about 3 times higher than 
is typical of very hot clusters $Z_{Fe} \simeq 0.25 Z_\odot$. 
}
{We have started a study of the metal distribution in nearby X--ray luminous poor galaxy clusters 
in the temperature range 3-4 keV with measured high abundances.
Our study aims at spatially resolving the metal content of the central regions of the ICM, in particular 
characterizing how our measurements are biased by the diagnostics adopted and by the possible
multi-temperature nature of the projected observed spectra. We report here on a 42ks \xmm\ observation
of the first object in the sample, the cluster \object{Abell 2028}.
 } 
{We selected interesting regions of the ICM to analyze the spatially
resolved structure of projected temperature and abundance, thereby producing two-dimensional maps.
}
{The \xmm\ observation of the first object of the sample,
the cluster Abell 2028, reveals the complex structure of the cluster over a scale of $\sim300$
kpc, showing an interaction between two subclusters in cometary-like configurations.
Cold fronts have been detected at the leading edges of the two substructures.
The core of the main subcluster is likely hosting a cool corona.
We show that a one-component fit for this region returns a biased high
metallicity. The inverse iron bias is caused by the behavior of the fitting code in
shaping the Fe-L complex. In the presence of a multi-temperature
structure of the ICM, the best-fit metallicity turns out to be
artificially higher when the projected spectrum is modeled with a
single temperature component, while it is not related to the presence
of both Fe-L and Fe-K emission lines in the spectrum. After
accounting for the inverse iron bias, the overall abundance of the
cluster is consistent with the one typical of hotter, more massive
clusters.
}
{We caution against interpreting high abundances inferred when fitting a single thermal component to spectra 
derived from relatively large apertures in 3-4 keV clusters, because the inverse iron bias can be present. 
Most of the inferences trying to relate high abundances in 3-4 keV clusters to fundamental physical processes
will probably have to be revised. 
}


\maketitle


\section{Introduction}

Clusters of galaxies have established themselves as a unique
environment for measuring elemental abundances and for studying 
the chemical enrichment history of the Universe. Their large
potential well retains all the metals produced by their member
galaxies, allowing in principle a unified description of the
thermodynamical properties of the hot phase (the diffuse X-ray
emitting gas) and of the cold phase (the cluster galaxies). Indeed,
the distribution of metals in the ICM is a direct consequence of the
past history of star formation in the cluster galaxies and of the
processes responsible for injecting enriched gas into the ICM.
\citep[For recent reviews of the observational and theoretical aspects, 
see][and references therein.]{Werner.ea:08,Borgani.ea:08,Schindler.ea:08}

Most of the appeal of the X-rays determining elemental abundances in the
hot gas stems from the apparent robustness of the measurement: most 
of the observed emission lines arise from well understood hydrogen- and
helium- like ions and their equivalent widths can be, under the reasonable
assumption of collisional equilibrium, directly converted into the elemental
abundance of the corresponding element. 
If some concerns arose for the 
reliability of the Fe-L shell modeling \citep[e.g.][]{Renzini:97,Arimoto.ea:97}, it has been
shown that this is not a particular issue \citep{Buote.ea:98,Buote.ea:03*1}

Unfortunately this intrinsic simplicity of the measurement has faced the
limitations of the X-ray satellites in terms of the shape of the instrumental
response, bandpass, and spectral and spatial resolution leading to observational biases. 
Correct modeling of the temperature structure is crucial, in
particular when dealing with spectra extracted from a large aperture
centered on the core of clusters and groups. Given the presence of
strong and opposite gradients in the temperature and metallicity
profiles \citep[i.e., cooler regions are more metal rich, see][]{De-Grandi.ea:01}, 
in the spectra of central regions photons coming from
regions of different temperature and abundances are mixed together.
These objects and these regions naturally draw the attention of X-ray observers because of their high
surface brightness, thus delivering high signal-to-noise (S/N) spectra.

The first important recognition of a bias in the measurement of elemental
abundances was the description of the ``iron bias" \citep{Buote.ea:98,Buote:00,Buote:00*1}.
Many \asca\ studies found significant subsolar values 
for the iron abundance in groups and elliptical galaxies 
\citep[e.g.][]{Davis.ea:99,Matsumoto.ea:97}, generally less 
than the stellar values in these systems and less than those observed in 
galaxy clusters. \citet{Buote:00*1} demonstrated that this was a bias resulting
from fitting with a simple single-temperature a multi-temperature plasma, 
resulting in best-fitting elemental abundances that are {\emph{too low}}.
The typical \asca\ spectrum for these objects was extracted from a large aperture in the core because of the 
large point spread function (PSF) of its telescope.
Since the description of this bias, and with the help of the
quantum leap increase in the quality of data with \xmm\ and \chandra, this kind
of error has been recognized and made less severe. The best-fitted abundances in the
multi-phase cooling core regions of clusters and in groups have been shown to 
increase when two-temperature models or more complicated DEM models are used 
\citep[e.g.][]{Molendi.ea:01,Buote.ea:03*1,Humphrey.ea:06*1,Werner.ea:06,Matsushita.ea:07}.
These new determinations helped to mitigate some of the trend seen in a plot
of iron-mass-to-light ratios (IMLR) versus cluster temperature,
which seem to drop by almost three orders of magnitude below 1~keV \citep{Renzini:97}.

Another trend in the ICM abundance versus cluster temperature has been shown with
increasing evidence in recent years. Using 
the \asca\ archive observations of 273 objects \citet{Baumgartner.ea:05}
showed that clusters with gas temperature between 2 and 4 keV have a typical mean 
abundance that is larger by up to a factor of 3 than hotter systems. 
Investigating a sample of 56 clusters with \chandra\ 
at $z\gtrsim0.3$, \citet{Balestra.ea:07}
confirms the trend of the Fe abundance with cluster temperature also in the
higher redshift clusters. 
The observed trend  may reflect a more efficient star formation in smaller clusters, 
as also suggested by optical and near--infrared observations of nearby systems 
\citep{Lin.ea:03,Gonzalez.ea:07}.
On the contrary it may be argued that it is difficult to conceive a physical mechanism
responsible for making clusters in the 3-4 keV range so unique a mass scale for chemical
enrichment \citep[see for example discussions in][]{Renzini:97,Renzini:04} and that some plausible measurement bias might be in action, 
which has to work in the
opposite way to the know Fe bias; i.e, it has to bias iron abundances {\emph{high}}.
A fundamental step has been made in this direction by \citet{Rasia.ea:08}, who analyzed
mock \xmm\ observations of simulated galaxy clusters finding a systematic overestimate of iron
for systems in the 2-3 keV range. The reason they put forward for this discrepancy is the 
complex temperature structure of the simulated clusters, due to projection and low spatial resolution, 
and to their particular temperature range.
\citet{Simionescu.ea:09} support this explanation by analyzing a deep \xmm\ observation of the
high-luminosity cluster Hydra~A finding a biased high Fe abundance at a level of 35\% in the
central three arcminutes, for the first time dubbing this kind of bias as the ``inverse'' Fe bias.
This appealing solution has been suggested by \citet{Rasia.ea:08} as a possible explanation
of the \citet{Baumgartner.ea:05} results.

With the aim of shedding more light on these issues by going beyond single-aperture measurements, 
we started with the present work a study of the metal distribution in iron-rich poor galaxy 
clusters ($3< kT_{\rm gas}<5$ keV) at redshift below $0.1$. From the work of 
\citet{Baumgartner.ea:05}, we selected the objects whose measurements lie at the highest end of the metal 
distribution. Here we present the analysis of the \xmm\ observation of Abell~2028, the first object in the sample.

Abell~2028 is a Bautz-Morgan type II-III galaxy cluster of Richness Class 1
located at redshift 0.0777 \citep{Struble.ea:99}. At this redshift 1 arcmin corresponds to 88 kpc.
The Galactic absorption in the cluster's direction is  
$2.46 \times 10^{20}$ cm$^{-2}$ \citep{Kalberla.ea:05}. 
\citet{Baumgartner.ea:05} used \asca\ to measure within an aperture of 7\arcmin\ an emission-weighted temperature 
of $3.91\pm0.61$ keV and metallicity of $0.77^{+1.25}_{-0.51} Z_{\odot}$ (error at the 90\% c.l.).

The outline of our work is the following.  
In Section~2, we describe the observation and data reduction. In Section~3
we describe the morphology of the cluster and discuss the surface brightness profiles
in selected regions. In Section~4 we discuss the spectral analysis and the temperature 
and abundances maps, highlighting the relevance of the inverse iron bias in the
central region of A~2028. We discuss the results of our analysis in Section~5. 
We summarize our results and draw conclusions for the present study
in Section~6. We show the results of simulations aiming at characterizing
the inverse iron bias in the Appendix.
Throughout this work, if not otherwise stated, we plot and tabulate values
with errors quoted at the 68.3 per cent ($1 \sigma$) level of confidence,
assuming a Hubble constant $H_0= 70 h_{70}^{-1}$ km s$^{-1}$ Mpc$^{-1}$ 
and $\Omega_{\rm m}=1-\Omega_{\Lambda}=0.3$.
All the metallicity estimates refer to the solar abundance in 
Anders \& Grevesse (1989) just for ease of comparison with previous work, because
this set of solar abundances is now outdated.

\begin{figure}
 \epsfig{figure=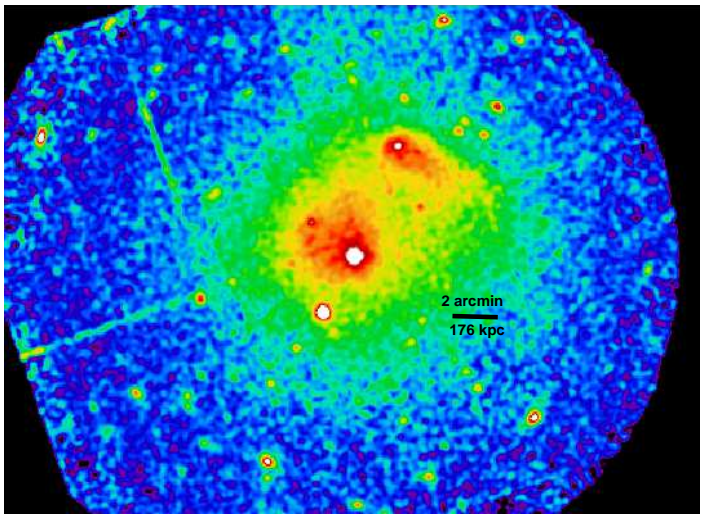,width=0.5\textwidth} 
\caption{Mosaic of the MOS1, MOS2, and pn images in the 0.5-2 keV energy band smoothed on a 
12'' scale. The image has been divided by the summed exposure maps to correct for exposure variations.
} \label{fig:ima}
\end{figure}

\section{Observation and data reduction}

The cluster Abell 2028 was observed by \xmm\ 
on July 29, 2007. The data were reduced with SAS v8.0 using the tasks
{\em emchain} and {\em epchain}. We only considered event patterns
0-12 for MOS and 0 for pn, and the data were cleaned using the standard
procedures for bright pixels and hot columns removal (by applying the
expression FLAG == 0) and pn out-of-time correction. The energy scale of the pn 
over the whole spectral bandpass was further improved with the task
{\em epreject}.
Periods of high backgrounds due to soft protons were filtered using a
threshold cut method by means of a Gaussian fit to the peak of the
histogram of the time bins of the light curve \citep[see Appendix
A of][]{Pratt.ea:02,De-Luca.ea:04} and excludede where the count rate
lies more than 3$\sigma$ away from the mean. The light curves were
extracted from regions of least source contamination (excising the
bright object core in the central 8\arcmin\ and the point source list
from the SOC pipeline, after visual inspection) in two different
energy bands: a hard band, 10-12 keV for MOS and 10-13 keV for pn (using 100s bins), 
and a wider band, 0.5-10 keV (using 10s bins), as a safety check for possible flares with
soft spectra \citep{Nevalainen.ea:05,Pradas.ea:05}. The flaring periods thus
determined were further checked by visual inspection of the light curves, resulting in
net exposures of 40.2 ks, 38.9 ks, and 29.5 for respectively the MOS1, MOS2, and pn detectors.
Point sources were detected using the task {\em ewavelet} in the energy band 0.5-10
keV and checked by eye on images generated for each detector. Detected
point sources from all detectors were merged, and the events in the
corresponding regions were removed from the event list, using circular
regions of 25\arcsec\ radius centered at the source position. The area
lost by point source exclusion, CCD gaps, and bad pixels was
calculated using a mask image. Redistribution matrix files (RMFs) and
ancillary response files (ARFs) were generated with the SAS tasks
{\em rmfgen} and {\em arfgen}, the latter in extended source
mode. Appropriate flux-weighting was performed for RMFs, using our own
dedicated software, and for ARFs using exposure-corrected images of
the source as detector maps (with pixel size of 1\arcmin, the minimum
scale modeled by {\em arfgen}) to sample the variation in emission,
following the prescription of \citet{Saxton.ea:02}.
For the background subtraction, we adopted a complete modeling of the various
background components as described in detail in \citet{Gastaldello.ea:07}.
We accurately model the actual sky background using a region free of cluster emission
at $r>10$\arcmin\ to the east where the signal-to-noise ratio of the surface brightness profile 
in the 0.5-2 keV band falls below 2.
Abell 2028 is projected towards a region of enhanced Galactic background, and in agreement
with previous studies \citep[see for example the case of Abell 2029, a cluster 
in the same region of the sky in][]{Vikhlinin.ea:05}, we found it necessary to add a third
thermal component at $\sim0.4$ keV. We checked the observation for contamination by solar wind 
charge exchange. The ACE SWEPAM\footnote{the data available at http://www.srl.caltech.edu/ACE/ACS/} 
proton flux was less than $4\times10^{8}$ protons s$^{-1}$ cm$^{-2}$
which is typical of the quiescent Sun \citep[e.g.,][]{Snowden.ea:04,Fujimoto.ea:07,Yoshino.ea:09}, and the 
light curve 
accumulated in a 0.5-2 keV energy band (after soft proton flares removal) did not show any variability. 
The spectra of the out-of-field-of-view events of CCD 4 and 5 of MOS1 and
CCD 5 of MOS 2 showed an anomalously high flux in the soft band \citep[see][]{Kuntz.ea:08}
so they were therefore excluded from our analysis.

\begin{figure*}[th]
\centerline{
\parbox{0.33\textwidth}{
\psfig{figure=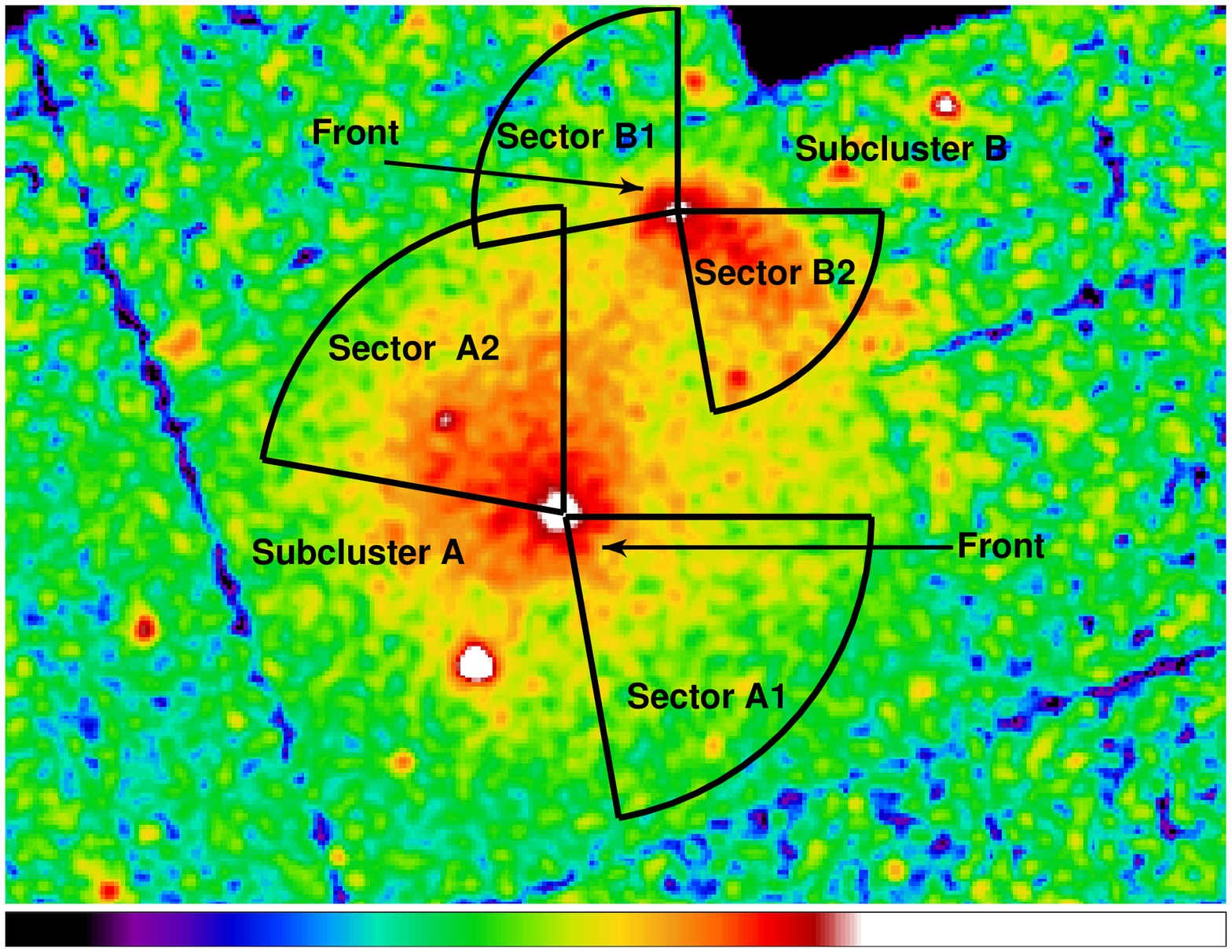,height=0.20\textheight}}
\parbox{0.33\textwidth}{
\psfig{figure=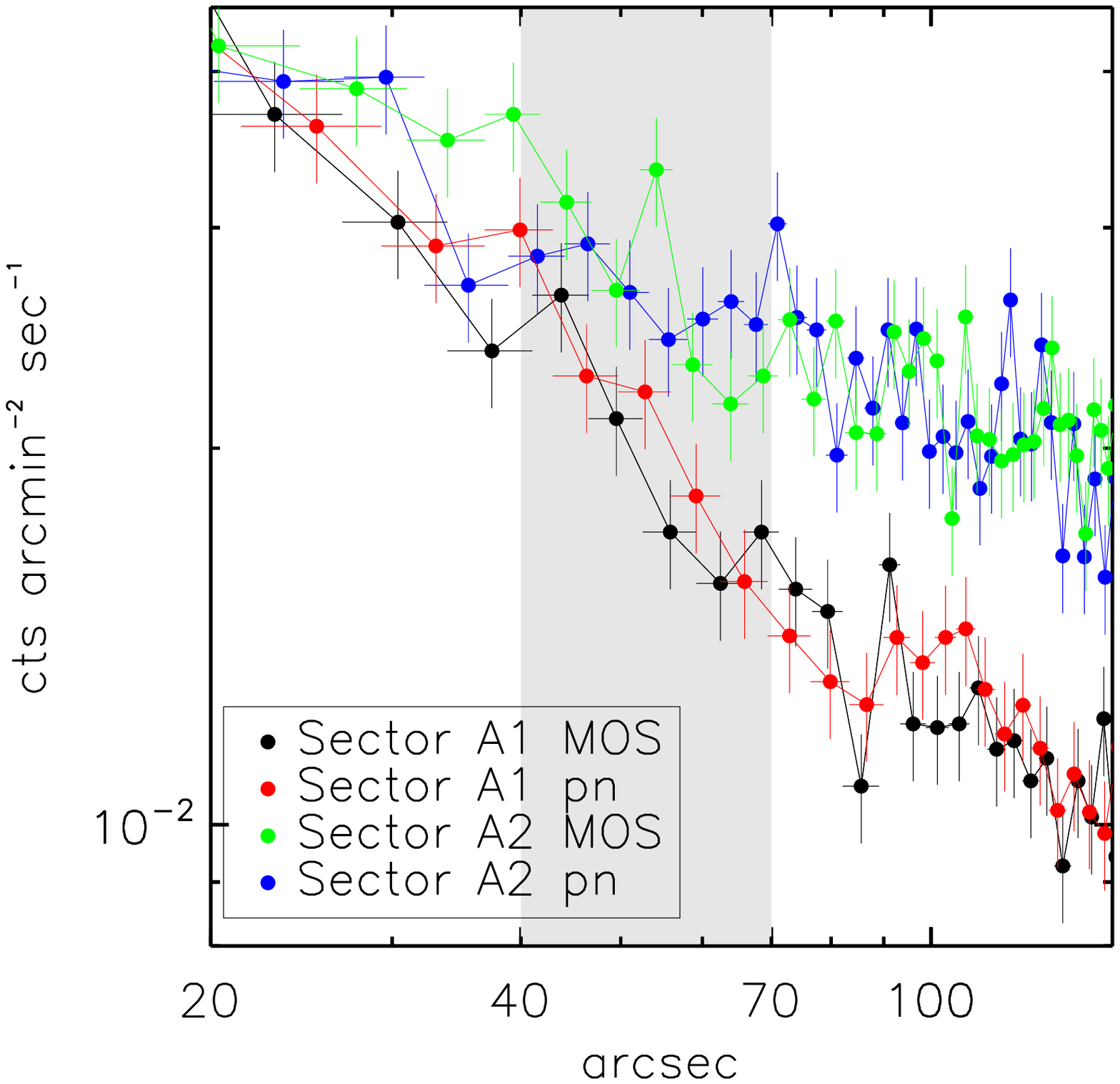,height=0.25\textheight}}
\parbox{0.33\textwidth}{
\psfig{figure=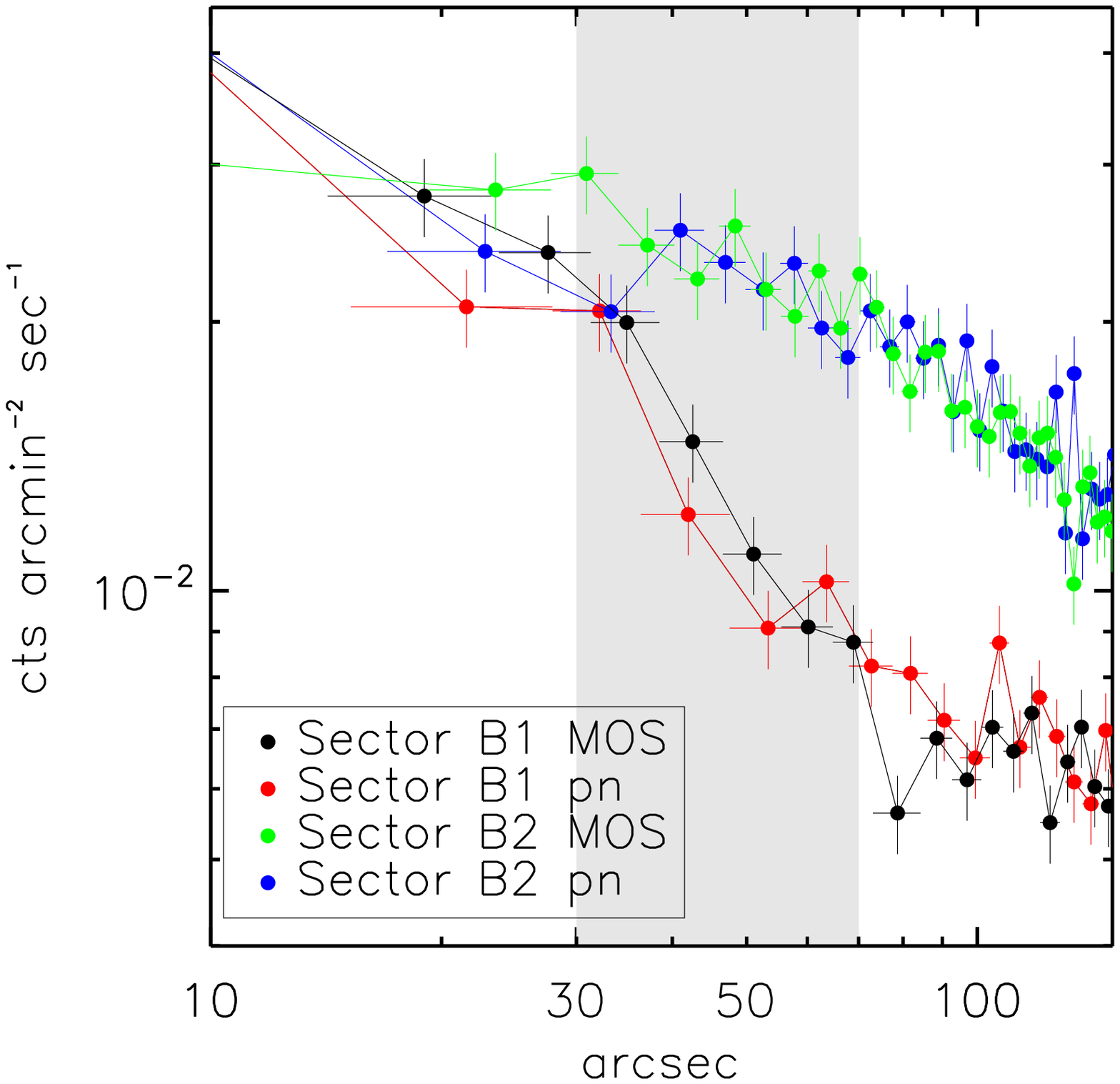,height=0.25\textheight}}
}
\caption{\label{fig:sectors} \footnotesize
\emph{Left panel:} Exposure-corrected MOS image in the 0.5-2 keV band with the angular sectors
used to extract the surface brightness profiles shown in the other panels. Point sources were masked
when extracting the profiles. The positions of the surface brightness jumps discussed in the text and shown in the
next panel are indicated by arrows.
\emph{Center panel:} Surface brightness profiles for the selected angular sectors shown in the left panel
for the main subcluster A. The radial ranges where surface brightness profiles change rapidly because of the surface brightness jumps 
are indicated by the shaded gray area.
\emph{Right panel:} same as the center panel, but for the smaller subcluster B.
}
\end{figure*}

\section{X-ray image and surface brightness profiles}\label{sec:surbri}

For each detector, we created images in the 0.5-2 keV band and exposure maps at the 
monochromatic energy of 1 keV and then combined the images into a single exposure-corrected 
image, smoothed on a scale of 12'', as shown in Fig.\ref{fig:ima}. The X-ray emission clearly 
reveals two merging subclusters, a more luminous SE component (which we will call subcluster A in the following
and labeled accordingly in Fig.\ref{fig:sectors}) and a less luminous 
NW one (subcluster B in the rest of the paper and labeled accordingly in Fig.\ref{fig:sectors}). 
The ``cometary'' shape of the two subclusters seems to suggest the direction
of the merging, with the subcluster A coming from the NE direction and the subcluster B coming from SW.
A further confirmation of this interpretation comes from the presence of surface brightness jumps
in the suggested direction of motion of the two subclumps: we extracted from the combined MOS and pn
exposure-corrected images profiles from selected angular sectors as depicted in Fig.\ref{fig:sectors}
centered on the surface brightness peaks of the two subclusters, with bins accumulated to have
100 counts. The surface brightness jumps are visible in the SW direction for subcluster A 
(sector A1, center panel of Fig.\ref{fig:sectors}) in the radial range 40\arcsec-70\arcsec\
and in the NE direction for the subcluster B (sector B1, right panel of Fig.\ref{fig:sectors}) in the radial 
range 30\arcsec-70\arcsec.
When coupled with the spectral information (see Section \ref{sec:1T}), these surface brightness jumps
can be interpreted as merging cold fronts.
For comparison plots of the profile in the direction of the tails of the two subclusters
are also shown (sector A2 for subcluster A and sector B2 for subcluster B), 
highlighting the clear asymmetry in the surface brightness distribution of
the two subclusters.

\begin{figure}
\centerline{
 \epsfig{figure=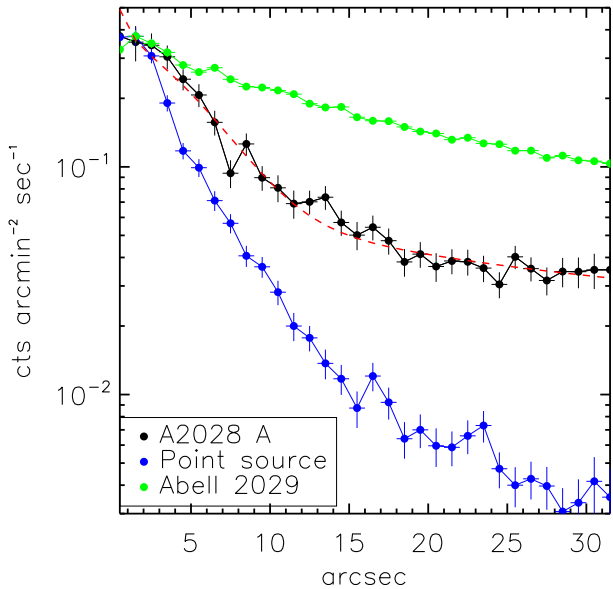,width=0.5\textwidth} }
\caption{Surface brightness profile of the core of subcluster A of A2028 compared to the
one of the bright point source SE of the subcluster and to the A2029 cluster, one of the clusters
with the most peaked surface brightness profile. The profiles have been scaled to match the peak
of the radial profile of A2028 A: the point source profile was multiplied by 0.29 and the
A2029 profile by 0.058. The dotted red line corresponds to the best-fit Gaussian+power-law model 
of the profile of the core of A 2028 A discussed in the text.} \label{fig:point}
\end{figure}

The X-ray peak location of the main subcluster A is consistent with the position of its brightest 
central galaxy, \object{GIN 416}. In Fig.\ref{fig:point} we show the surface brightness profiles extracted
from the combined MOS image and accumulated in bins of 1.1\arcsec\ (corresponding to the MOS
physical pixel size; MOS is better suited to this type of analysis because it oversamples the core of the PSF,
on the contrary the pn pixel physical size is 4.1\arcsec). In the same figure we plot the profile of 
the bright point source (RA 15:09:34, DEC +07:30:54) SE of the main subcluster and the \xmm\ profile of 
\object{Abell 2029} (OBSID 0111270201), one of the clusters with the most peaked surface brightness profile and at the same 
redshift of A2028. Profiles are scaled to match their peaks of 
surface brightness. 
In the innermost 8\arcsec, the surface brightness data indicate of an excess. We fitted the inner 30\arcsec\ of
the A 2028 A profile with a Gaussian plus a power law, the latter component modeling the ICM emission. The 
best fit model, with a $\chi^{2}/\rm{dof}=29/28$ gives a width for the 
Gaussian of $\sigma = (4.9\pm0.3)$\arcsec, which corresponds to an FWHM
of $(11.5\pm0.7)$\arcsec\ bigger than the on-axis FWHM of the \xmm\ PSF, which is 4.4\arcsec\ for MOS at 1.5 keV.
\footnote{\xmm\ Users Handbook, http://xmm.esac.esa.int/\-external/xmm\_user\_support/documentation/uhb/node17.html}
This excess central component, which is more extended than a point source, may be a corona associated 
with the BCG, as seen in many merging clusters \citep{Vikhlinin.ea:01*1,Sun.ea:07}. Coronae are sources extended on 
kpc scales, and they are the remnants of galactic X-ray coronae with temperatures around 1 keV, compressed by the 
hot gas of the cluster.
The spectral information we present in Section \ref{sec.core} will strengthen this interpretation.

\section{Spectral analysis}

\subsection{Temperature and abundance maps}
\label{sec:1T}

\begin{figure*}[th]
\centerline{
\parbox{0.5\textwidth}{
\psfig{figure=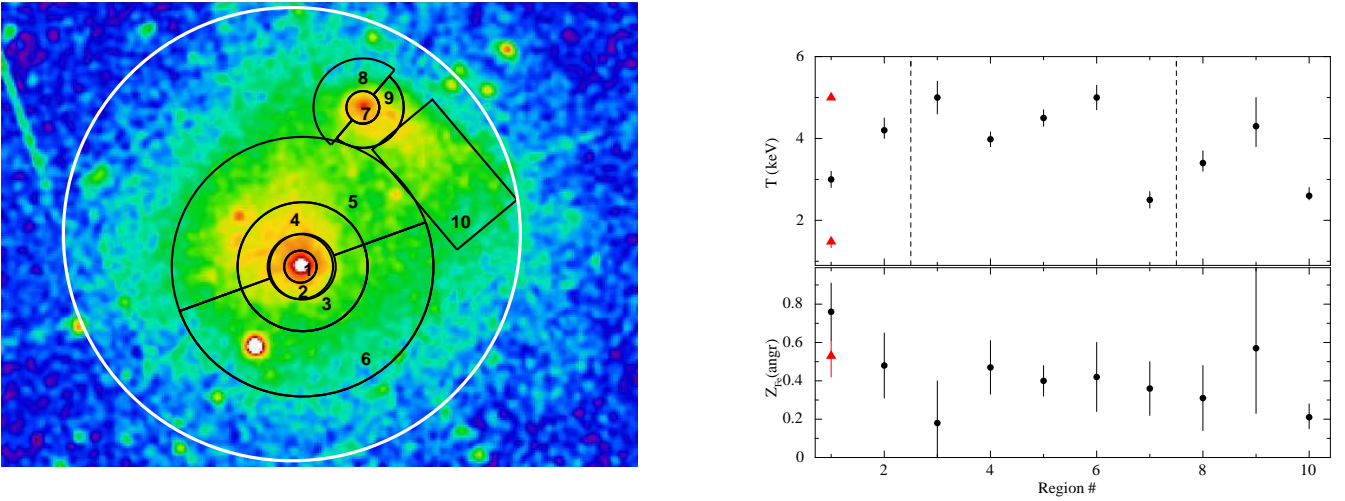,height=0.25\textheight}}
\parbox{0.5\textwidth}{
\psfig{figure=values_map_paper_new.ps,angle=-90,width=0.45\textwidth}}
}
\caption{\label{fig.tmap} \footnotesize
\emph{Left panel:} Regions used for spectral extraction to derive the temperature and abundance maps. 
The white circle corresponds to the \asca\ aperture. 
\emph{Right panel:} Black circles correspond to the values and $1\sigma$ error bars for the temperature and abundances 
obtained with the 1T model for each region shown in the left panel, discussed in the text in Section \ref{sec:1T}. 
The dashed vertical lines indicate the regions adjacent to the surface brightness jumps of Figure \ref{fig:sectors}.
The red triangles correspond to the values for temperature and abundance obtained with a 2T model, as discussed in the 
text in Section \ref{sec.core}. The higher temperature value at 5 keV has been fixed.
}
\end{figure*}

We extracted spectra in a series of regions as depicted in the left panel of 
Fig.\ref{fig.tmap}, chosen to select relevant regions of the ICM.
Given the complex geometry and the relatively low number of counts we did not de-project
the spectra and the results quoted in this paper come from fitting projected spectra.
They consist in two concentric annular regions centered on the emission peak of the
subcluster A (RA 15:09:28 DEC +07:33:20) with bounding radii 0\arcmin-0.5\arcmin, 
0.5\arcmin-1\arcmin\ and four sectors with bounding radii 1\arcmin-2\arcmin\ and 2\arcmin-4\arcmin; 
a circle of 0.5\arcmin\ radius centered on the emission peak of the subcluster B (RA 15:09:20 DEC +07:38:15), 
two sectors with  bounding radii 0.5\arcmin-1\arcmin\
and a box of 4\arcmin$\times$2.4\arcmin\ encompassing the tail.
The radii of regions 2 and 3 and of regions 7 and 8 have been determined to match 
the surface brightness jumps discussed in Section \ref{sec:surbri}
as closely as possible, in order to investigate 
the temperature variation across these fronts.
We fitted the background subtracted spectra with a single APEC thermal plasma model
\citep{Smith.ea:01} in the 0.5-10 keV band with the absorbing column density fixed at the 
Galactic value. The temperature and abundance values of the regions of the map thus 
obtained are shown in the right panel of Fig.\ref{fig.tmap}.
The temperature map reveals that subcluster A has a uniform temperature distribution in the range 4-5 keV 
with a drop in the inner 30\arcsec\ region to 3 keV, whereas subcluster B seems to 
have a temperature gradient, increasing from 2 keV in the surface brightness peak to 4 keV.
The abundances in every region chosen for spectral analysis show values in the
range 0.3-0.5 \solar, with the exception of the central 0.5\arcmin\ region of subcluster A. In this region a single
temperature fit returns a high metallicity of $0.76\pm0.14$ \solar.
The regions corresponding to the surface brightness
jumps described in Section \ref{sec:surbri} are consistent with being cold fronts, because
in subcluster A the temperature increases from $4.2^{+0.2}_{-0.1}$ keV in region 2 to
$5.0\pm0.4$ keV in region 3 and in the subcluster B from $2.5\pm0.2$ keV in region 7 to
$3.4^{+0.3}_{-0.2}$ keV in region 8.

\begin{figure*}[ht]
\centerline{
\parbox{0.33\textwidth}{
\psfig{figure=speccore_05_10_new.ps,angle=-90,width=0.33\textwidth}}
\parbox{0.33\textwidth}{
\psfig{figure=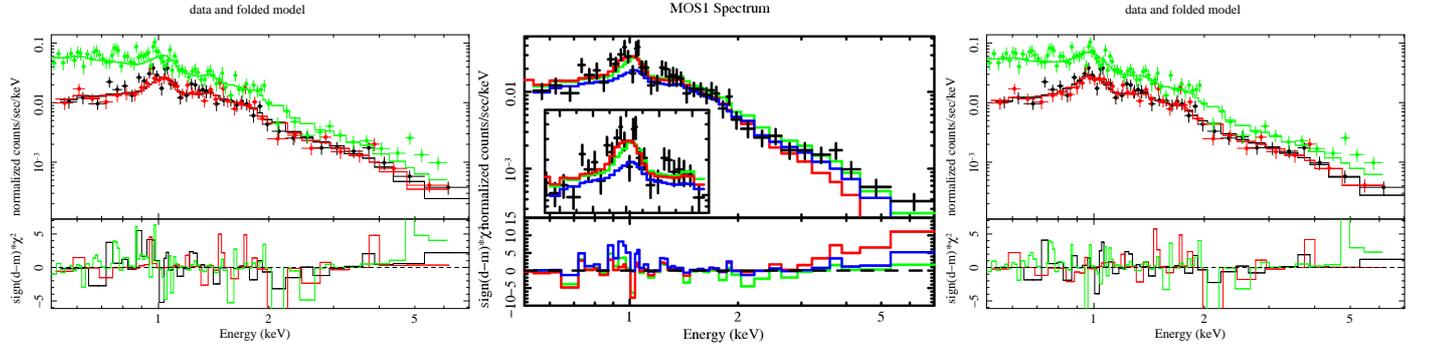,width=0.33\textwidth}}
\parbox{0.33\textwidth}{
\psfig{figure=speccore_05_10_2T.ps,angle=-90,width=0.33\textwidth}}
}
\caption{\label{fig.speccore} \footnotesize
\emph{Left panel:}  X-ray spectrum of the core of the main subcluster of Abell 2028
taken from a 0.5\arcmin\ aperture (corresponding to region 1 of Fig.\ref{fig.tmap}). Data
from MOS1, MOS2, and pn are plotted in black, red, and green, respectively. The
best-fit model obtained from a 1T fit in the broad band 0.5-10 keV and residuals are also shown.
\newline
\emph{Central Panel:} MOS 1 spectrum from the inner core corresponding to region 1.
Three different spectral models are overplotted: the 1T model (green line) fitted in the broad 0.5-10 keV band;
the 1T model fitted in the 0.5-3 keV band (red line) and the 1T model with the parameters obtained by 
the 0.5-10 keV fit, but with the abundance value kept fixed at the lower value obtained in the 0.5-3 keV band 
(blue line).
\newline
\emph{Right panel:} Same as in the left panel with the best-fit 2T model (with the higher temperature fixed at 5 keV) 
in the 0.5-10 keV band.
}
\end{figure*}

\subsection{The spectrum of the core of subcluster A}\label{sec.core}

We focus here on the spectrum of the core of the main subcluster, corresponding to region 1
in Fig.\ref{fig.tmap}, which is a circular region of 0.5\arcmin\ radius, corresponding to 44 kpc.
The single temperature fit performed in Section \ref{sec:1T} is a problematic fit because of 
residuals around the structure of the Fe-L shell,  as can be seen in the left panel of Fig.\ref{fig.speccore}, and
the high $\chi^{2}/\rm{dof}=243/168$ highlights it further (see Table \ref{tab:core} for the results
of all the fits discussed in the text for this region).
The Fe-K line complex is not evident in the data, and a fit in an energy
range (0.5-6.0 keV) which excludes the redshifted energy of the Fe-K line (6.2 keV) returns 
a metallicity of $Z=0.55\pm0.10$ \solar.
If we restrict the energy band in the 0.5-3 keV range in an attempt to better model
the Fe-L complex, we find a lower temperature ($1.99\pm0.09$ keV with respect to $2.95\pm0.13$ keV) and 
abundance ($Z=0.32^{+0.07}_{-0.05}$ \solar\ 
with respect to $Z=0.76\pm0.14$ \solar) compared to the broad band fit 
but with an improved fit, $\chi^{2}/\rm{dof}=178/148$.
Clearly this model underestimates the data at high energy when applied to the broad 0.5-10 keV band, as 
shown in the central panel of Fig.\ref{fig.speccore} where we show the MOS 1 spectrum with the best-fit 1T models 
in the broad and soft energy band overplotted with the shape of the Fe-L shell broad feature 
and residuals between the two fits in the inset. 

These results clearly indicate the need for a range of temperatures for fitting the spectrum
of the core of A2028 A and we therefore investigated fitting a two-temperature (2T)
model to the data with the abundances tied between the two thermal components because the 
quality of the data does not allow relaxing this condition. The 2T fit 
does not constrain the temperature of the hotter component and the results  
of the fits obtained by fixing the temperature of the hotter component in the range 5-6 keV, as suggested by the 
temperature values in the regions of Fig.\ref{fig.tmap}, are listed in Table \ref{tab:core}. Using a temperature 
of 6 keV or 5 keV for the hot component leads to statistically identical fits, with a temperature for the cooler 
component in the range
1.20-1.42 keV and metallicity in the range 0.53-0.75 \solar. The 2T model is a better description of the data
with $\chi^{2}/\rm{dof}=204/167$ corresponding to a significance of the additional component, calibrating 
the $F$-statistic using simulations of the null model (the 1T model fit above) as suggested 
by \citet{Protassov.ea:02}, of 99.99 \%.

We investigated other multi-temperature models to attempt to constrain the general differential emission 
measure (DEM) of the emission spectrum; in particular, following \citet{Buote.ea:03}, we used a model in 
which the temperature distributions is a Gaussian (GDEM) centered on the mean $T_0$ with width $\sigma_T$ 
(one additional parameter compared to the 1T model) and a model where the temperature distribution is a 
power law (PDEM) with slope $\alpha$ and width of the temperature distribution characterized 
by $T_{max}$ and $T_{min}$ 
(two additional parameters compared to the 1T model). The abundances are tied between all the thermal components.
The fits with these multi-temperature models are comparable to the one obtained with the 2T model, and they confirm
the composite nature of the spectrum of the core. We also performed the fits using the MEKAL plasma code 
\citep{Mewe.ea:85,Liedahl.ea:95} finding similar results.

\begin{figure}
\centerline{
 \epsfig{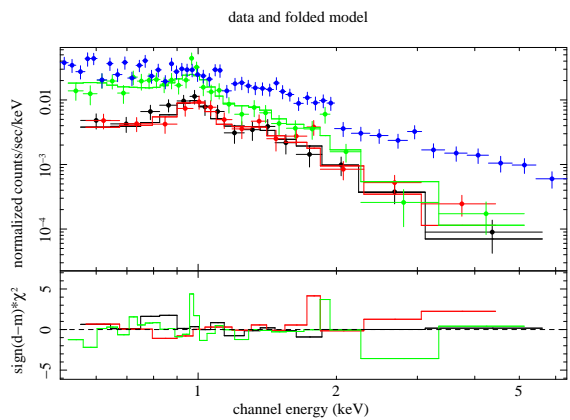} }
\caption{ X-ray spectrum of the inner 10\arcsec\ of the core of Abell 2028 A (having subtracted the emission coming from
an annulus with bounding radii 15\arcsec-30\arcsec). Data
from MOS1, MOS2, and pn are plotted in black, red, and green, respectively. The
best-fit model obtained from a 1T fit and residuals are also shown.
The pn spectrum taken from the 15\arcsec-30\arcsec\ annulus is shown in blue 
for comparison.
} \label{fig:speccorona}
\end{figure}

Given the evidence of a soft component with temperature in the range 1.3-1.5 keV and the surface 
brightness profile shown in Fig.\ref{fig:point}, we made an attempt (bearing in mind all the caveats of 
extracting a region 
comparable in size with the half energy width of the PSF) to extract a spectrum from a circular region of 
10\arcsec\ radius using as background a spectrum extracted from an annulus with bounding radii 15\arcsec-30\arcsec, 
both centered on the emission peak. The purpose of this attempt is to investigate the excess emission that might 
be associated with the presence of a cool corona. The use of the local background very close to the excess emission
should subtract the hot thermal emission of the surrounding ICM thereby revealing the cooler emission of the corona.
The spectrum of the inner 10\arcsec\ is well fitted by a 1T model with $kT = 1.46\pm0.10$ keV. 
We find the same result if we simultaneously fit the two spectra and we take into account the spreading 
of the cooler X-rays from the very center due to the finite PSF by means
of the cross-talk modification of the ARF generation software \citep[e.g.][]{Snowden.ea:08}.
The spectrum with the best-fit model is shown in Fig.\ref{fig:speccorona} together with the pn spectrum of the 
15\arcsec-30\arcsec\ annulus to represent the hotter ICM in which the corona is embedded.

\begin{table*}[ht]
\caption{Results of the fits using single and multi-temperature models (see text for details of the models ) 
in different energy bands for the EPIC spectrum of the core (circle of 0.5\arcmin\ radius) of A 2028 A.}
\label{tab:core}
\begin{center}
\begin{tabular}{lccccccccl} \hline\hline\\[-6pt]
Model/Band & $kT_0$ &  Z  &  Norm$_0$ & $kT_1$ &  Norm$_1$ & $\sigma_{T}$/width & $\alpha$ & $\chi^2$/dof & Notes\\
\hspace{0.8cm} (keV)        & (keV)    &  ($Z_{\odot}$) & ($10^{-4}\rm{cm}^{-5}$) & (keV) & ($10^{-4}\rm{cm}^{-5}$)  & (keV) & & \\
\hline
1T 0.5-10    & $2.95\pm0.13$ & $0.76\pm0.14$  &  $1.70\pm0.09$                &                        &                        &  & & 243/168 & Full spectrum\\
1T 0.5-6     & $2.59\pm0.10$ & $0.55\pm0.10$  &  $1.93\pm0.09$                &                        &                        &  & & 234/165 & no Fe-K \\
1T 0.5-3     & $1.99\pm0.09$ & $0.32^{+0.07}_{-0.05}$  & $2.12\pm0.09$        &                        &                        &  & & 178/148 & soft band Fe-L\\
2T 0.5-10    & 5.0 (fixed) & $0.53_{-0.11}^{+0.08}$  & $1.20^{+0.11}_{-0.06}$ & $1.48_{-0.15}^{+0.07}$ & $0.65_{-0.18}^{+0.05}$ &  & & 204/167 & \\
2T 0.5-10    & 4.0 (fixed) & $0.75_{-0.08}^{+0.13}$  & $1.42^{+0.06}_{-0.03}$ & $1.27_{-0.03}^{+0.07}$ & $0.27_{-0.05}^{+0.02}$ &  & & 204/167 & \\
GDEM 0.5-10  & $3.92^{+0.16}_{-0.22}$ & $1.18^{+0.28}_{-0.26}$  & $1.69\pm0.13$        &                      & & $1.67^{+0.24}_{-0.22}$ & & 202/167 & \\ 
PDEM$^{1}$ 0.5-10  & $0.44\pm0.08$  & $1.55^{+0.45}_{-0.26}$  & $1.36\pm0.12$       &       & & $5.81^{+0.80}_{-0.71}$  & $0.83^{+0.21}_{-0.18}$ & 197/166 & \\  
        
\hline\\
\end{tabular}
\end{center}
$^{1}$ For PDEM $T_0$ corresponds to $T_{min}$ and the width to  $T_{max}-T_{min}$
\end{table*}

\subsection{The spectrum of the large \asca\ aperture}

We extracted a spectrum from a circular region of 7\arcmin\ centered on RA 15:09:29 DEC +07:34:20
\citep{Horner:01}, which corresponds to the \asca\ aperture (the white circle in Fig.\ref{fig.tmap}).
We fitted a 1T model and multi-temperature models as done for the core of A2028 A. As expected given the wide range of temperatures revealed by the 
temperature map of Fig.\ref{fig.tmap}, the 1T model fails to accurately model the spectra of this \asca\ aperture which 
include the bulk of emission of the two subclusters. The multi-temperature models also do not fit the spectra adequately.
But significantly we do not see the large variations in metallicity among different models as seen in the case
of the core of Abell 2028 A, and it is close to the average abundance in the various regions of this merging cluster
as revealed by the two-dimensional map. 
There are enough counts to allow us to fit the spectrum
just in the 2-10 keV energy band and to measure an abundance based just upon the Fe-K line. The results are 
consistent within the errors with all the previous determinations ($0.35\pm0.05$ \solar\ obtained with a 1T fit in the 2-10 keV band, 
to be compared with the upper value of the 
range, $0.43\pm0.03$ \solar\ obtained with the 1T and GDEM models in the 0.5-10 keV band, and the lower value, $0.33\pm0.04$ \solar\ 
obtained with a 2T model in the 0.5-10 keV band).

\begin{figure*}[t]
\centerline{
\parbox{0.5\textwidth}{
\psfig{figure=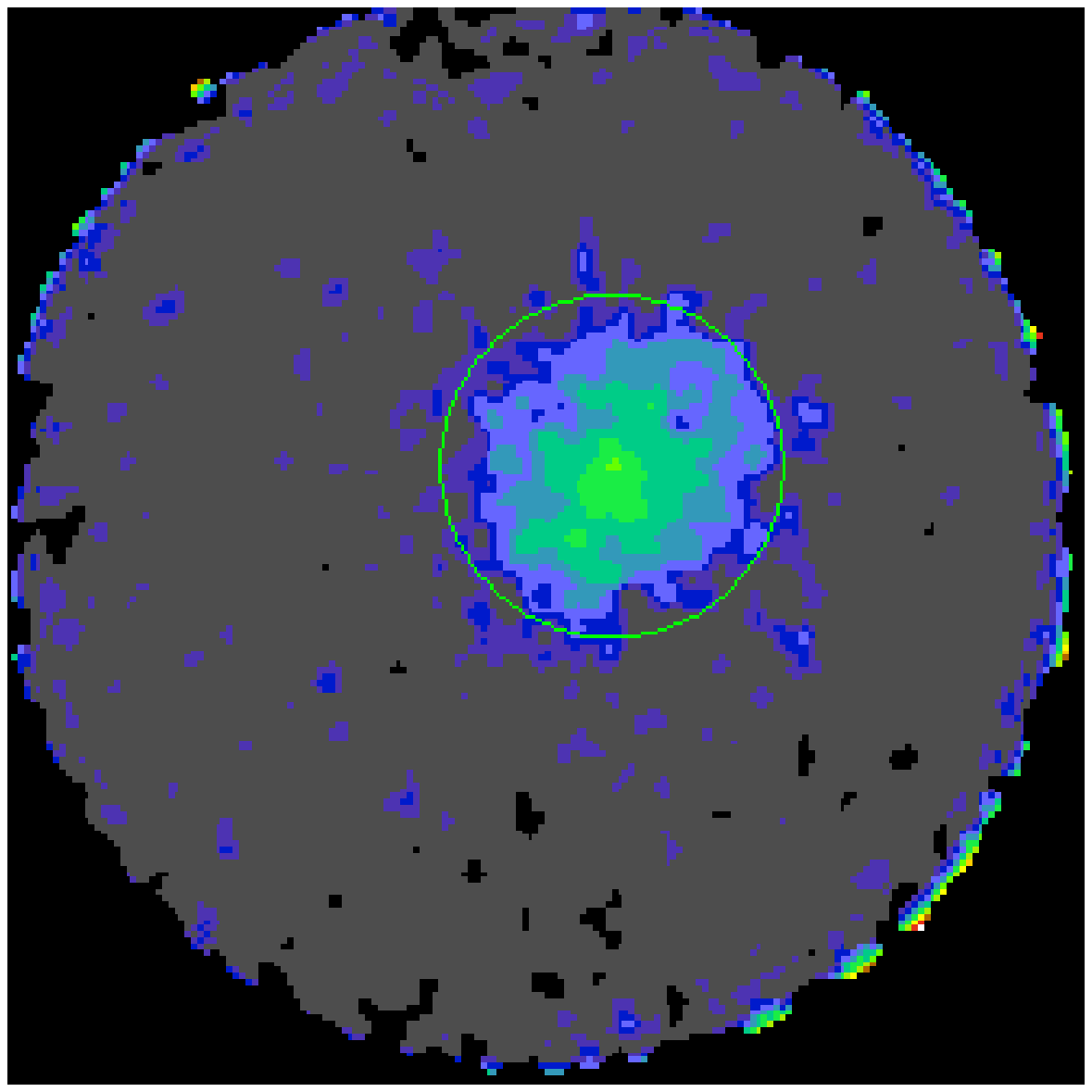,width=0.35\textwidth}}
\parbox{0.5\textwidth}{
\psfig{figure=comparison_asca.ps,angle=-90,width=0.4\textwidth}}
}
\caption{\label{fig:imageasca} \footnotesize
\emph{Left panel:}  Exposure-corrected \asca\ GIS image of Abell 2028. The regions inside the solid green
lines (R=7$^{\prime}$) were extracted for the source spectrum analyzed in Baumgartner et al. (2005).
Figure taken from http://asd.gsfc.nasa.gov/Donald.Horner/acc/html/ABELL\_2028\_83040050.html
\newline
\emph{Right panel:} Comparison of the X-ray spectra of the large \asca\ aperture 
from the \asca\ SIS 0 and 1 detectors (black and red, respectively) and from the EPIC pn (green).    
}
\end{figure*}

\begin{figure*}[ht]
\centerline{
\hspace{1.0truecm}
\parbox{0.5\textwidth}{
\psfig{figure=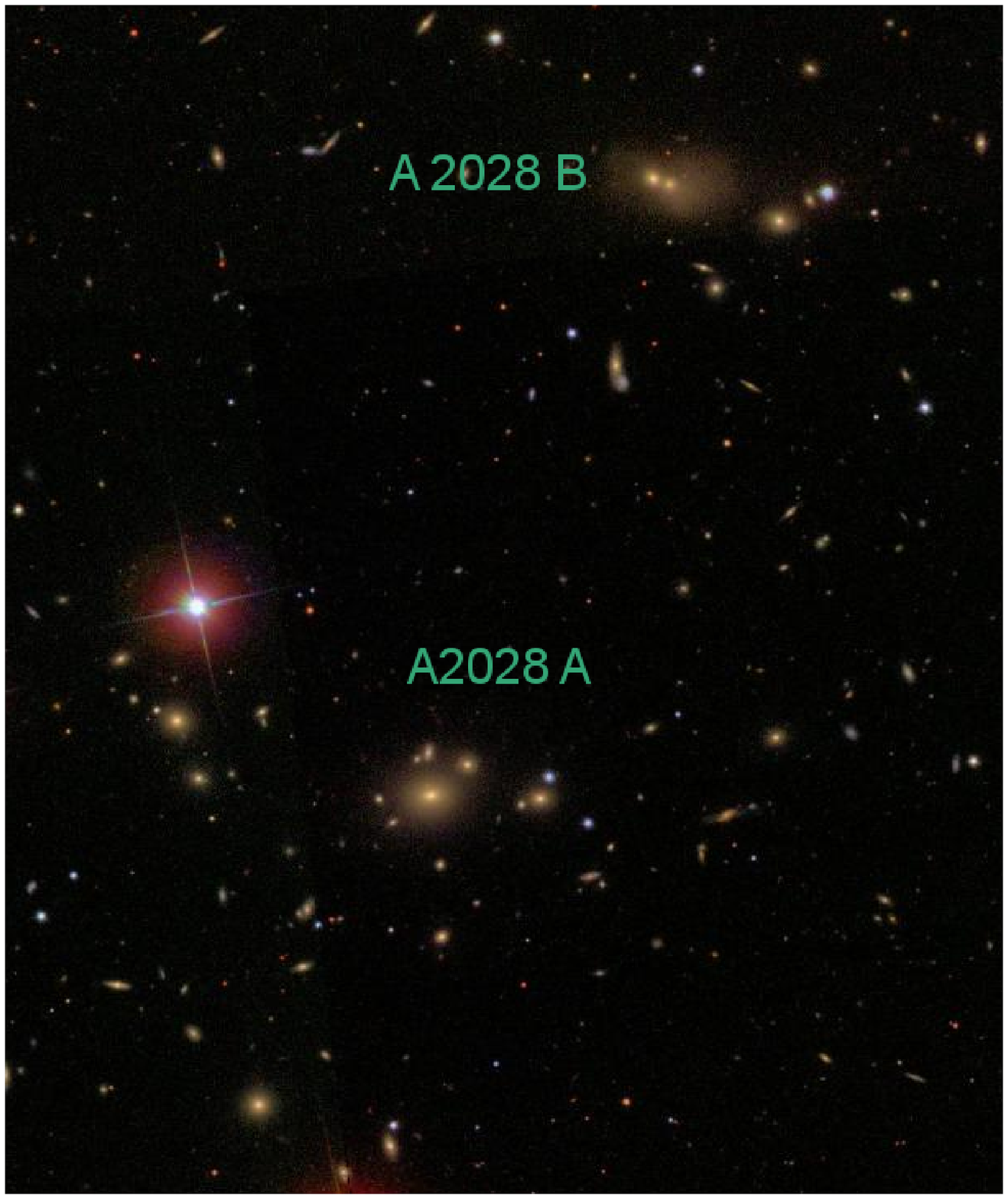,width=0.33\textwidth}}
\parbox{0.5\textwidth}{
\psfig{figure=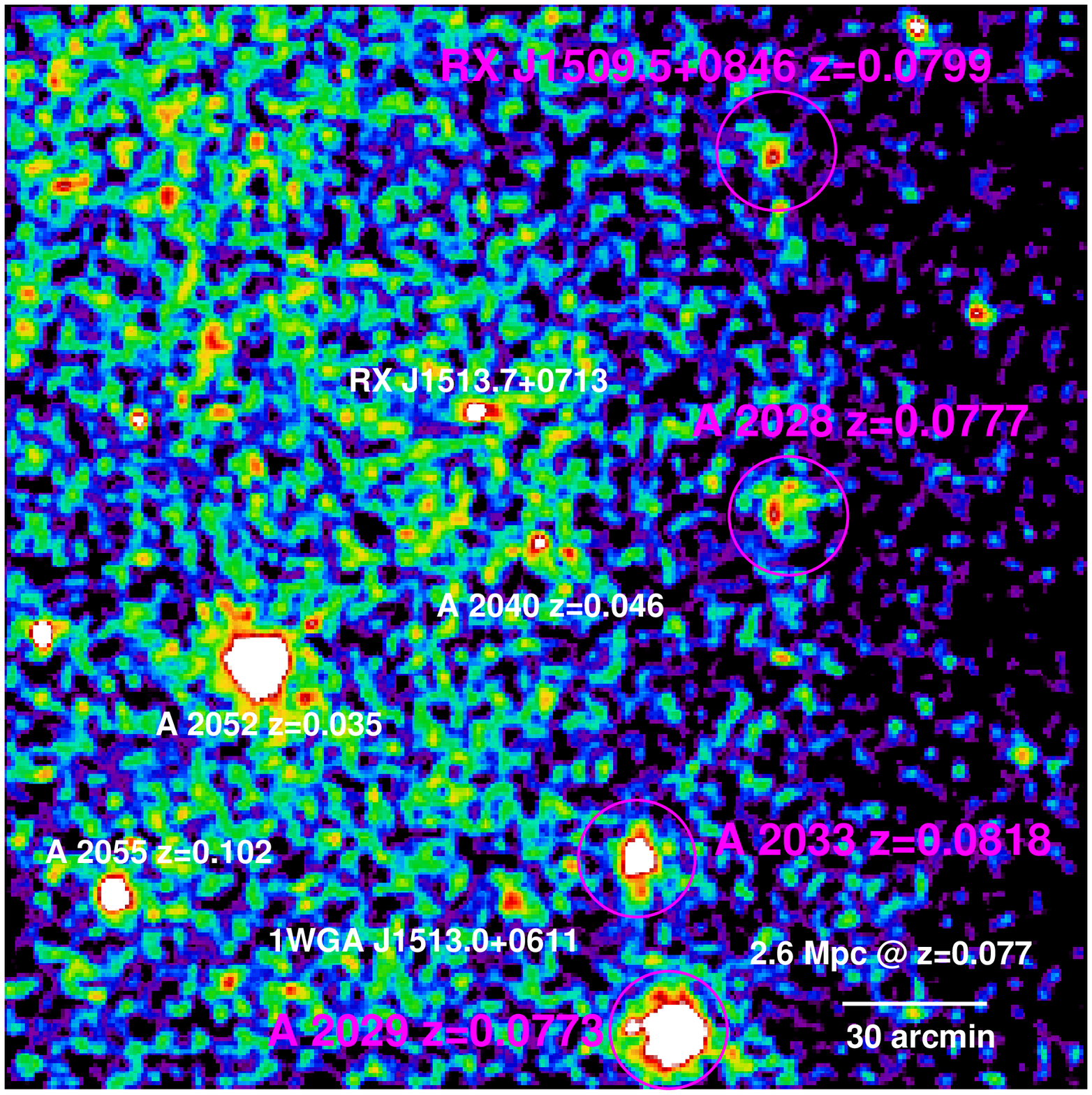,width=0.36\textwidth}}
}
\caption{\label{fig.optical} \footnotesize
\emph{Left panel:} SDSS color image of the field of Abell 2028
\emph{Right panel:} The RASS field of Abell 2028 highlighting in magenta the other members of 
the supercluster as Abell 2033 and Abell 2029.
}
\end{figure*}

\section{Discussion}

\subsection{The inverse iron bias at work in Abell 2028}

The \xmm\ observation of Abell 2028 clearly shows its complex structure by revealing
two merging subclusters, a more luminous SW component (A2028 A) and a less luminous 
NE component (A2028 B). Associated with the brightest central galaxy (BCG) of A2028 A,
GIN 416, there is indication of a cool corona as suggested by the excess 
surface brightness profile and a cooler spectral component \citep{Sun.ea:07}. The BCGs that able to retain
their corona are usually associated with merging clusters, such as the first observed objects of this class,
the BCGs in the Coma cluster \citep{Vikhlinin.ea:01*1}. 

It is therefore clear that, within the beam of the \asca\ region used to derive the abundances
presented for A2028 in the work of \citet{Baumgartner.ea:05}, several components with different temperatures
are present. Work by \citet{Mazzotta.ea:04} and \citet{Vikhlinin:06*1} pointed out how the temperature obtained
by a fit with a single-temperature plasma model can be significantly different from commonly used definitions
of average temperature like the emission-weighted temperature. 
However, as pointed out by \citet{Mazzotta.ea:04}, the spectrum of a
multi-temperature model cannot be reproduced correctly by a
single-temperature thermal model. On the other hand, a
multi-temperature thermal emission may be well-fitted by a
single-temperature model only because of the convolution with the
instrument response, Poisson noise and background noise.
Even more difficult to gauge is the concept of the spectroscopic-like
abundances, i.e. the measured abundance using a single temperature fit to a multi-temperature spectrum
\emph{with possible different abundances in different temperature components}, because it depends strongly
on the correct modeling of the temperature structure and on the limitations of current CCD detectors, which
cannot spectrally separate emission from different components in a precise way. Besides the early discovery
of the Fe bias \citep{Buote:00*1}, only recently work has been done to investigate this issue 
\citep{Mazzotta.ea:04,Balestra.ea:07,Rasia.ea:08}.

In the work by \citet{Rasia.ea:08}, who analyzed six simulated galaxy clusters processed through an 
X-Ray Map Simulator \citep[X-MAS,][]{Gardini.ea:04} that allows mock \xmm\ MOS 1 and MOS 2
observations to be created, a systematic overestimate of iron for systems in the 2-3 keV range was found.
By simulating a spectrum resulting from the combination of plasmas 
at 2 and 3 keV, they found fitted iron abundances higher with respect to the input value up 
to 40\% in low S/N spectra (but still with 4500 counts). They also found the very interesting fact
that the overestimate depends inversely on the number of counts, i.e. low S/N spectra have a greater bias.
Projection and low resolution effects can create a complex temperature structure, averaging  
different temperatures in the same radial bin. 
\citet{Simionescu.ea:09} give further support to this interpretation analyzing a 3\arcmin\ aperture
in the center of the cool core cluster Hydra A finding that a GDEM model is a better fit to the 
spectrum and that it returns a best-fit Fe abundance lower than the one obtained from a fit 
with a single-temperature model of the full spectral band. (Lower values are derived when fitting, in particular, 
just the soft band.) 

The statistical quality of either the simulated spectra by \citet{Rasia.ea:08} or the Hydra A spectra
presented in \citet{Simionescu.ea:09} are good, and in particular the Fe-K feature is clearly present. 
Many of the objects in the \asca\ sample did not have sufficient signal-to-noise spectra for the Fe-K line emission
to be present with the same statistical significance also because \citet{Baumgartner.ea:05} did not include very 
bright clusters to not unduly bias the results of their stacking analysis.
This is indeed the case for the \xmm\ spectrum of the core of A 2028 A or the \asca\ spectra of the 7\arcmin\ large aperture of the 
cluster (see right panel of Fig.\ref{fig:imageasca}). With the aid of the simulations presented in  
appendix \ref{sec.appendix}, we suggest that the inverse-Fe bias is mainly driven by the behavior of the the Fe-L complex 
strength which falls rapidly with increasing temperature above roughly 3 keV: in a multi-T spectrum where the
average T is near 3 keV, but there are temperature components below and above 3 keV, most of the the Fe
L emission of this multi-T spectrum comes from the lower-T components in the spectrum. However, 
when one fits a (wrong) single-T model, the single temperature will be sufficiently high that
the Fe L lines will be weaker, and thus the model will have to compensate by increasing the Fe abundance above the true value.
This is also shown by the central panel of Fig.\ref{fig.speccore}.
That the Fe-K line is not necessary for the inverse iron bias to work can also be demonstrated by 
simulating 1T spectrum with the characteristics of the fit in the 0.5-3 keV band for the spectrum shown in 
Section \ref{sec.core}, i.e. a single thermal spectrum with $kT = 2.0$ keV and $Z=0.38$ \solar. If we force the 
temperature in the model fit to be fixed at 3 keV (as in the case of a broad band fit where higher temperature 
components are present in the spectrum), the abundance will be biased high at a value of $Z=0.68$ \solar.
The thermal structure of a merging cluster is different from the very central regions of a cool core cluster
and the main driver of the inverse Fe-bias in A2028 A is the cool component of the corona, which also justifies
the simulations of Appendix \ref{sec.appendix} and fits with a 2T model, which seems a good physical description  
of a corona embedded in the hot intra-cluster gas. More complicated multi-temperature models, like the GDEM model, may 
provide on the contrary a more physical interpretation of the spectrum of the central regions of cool core clusters, as 
suggested by \citet{Simionescu.ea:09}.

The inverse-iron bias is, however, not present in the \xmm\ EPIC spectra of the large 7\arcmin\ aperture.
The S/N of this spectrum is higher and the presence of a cool component is less important. This reinforces the fact
already pointed out by \citet{Rasia.ea:08} that the bias depends inversely on the number of counts present in the spectrum.
It is beyond the scope of this paper to investigate the reasons for the anomalous abundances directly in 
the \asca\ spectra but the presence of the inverse iron-bias in those low S/N spectra (obtained with a lower effective 
area instrument) is a reasonable suggestion. The \xmm\ results for A2028 show that the abundances in this cluster are 
not anomalously high.

The presence of the direct iron bias \citep{Buote:00*1} and of the inverse iron bias described in this paper show the
danger of interpreting abundances measured with single-temperature fits
to spectra extracted from large physical apertures (and of relatively low S/N), where there might be a great 
deal of temperature structure. This is particularly true for the combination of current telescopes and CCD detectors,
which give a lot of weight to the soft energies and therefore to the shape of the Fe-L complex. Caution should be paid,
for example, when interpreting global abundances measured in clusters out to a redshift of 
$z\sim$0.5 because the Fe-L shell of low-temperature components
(down to $\sim$1 keV) is within fitting bandpasses with lower energy ends of 0.5-0.6 keV.

\begin{figure*}[ht]
\centerline{
\hspace{1.0truecm}
\parbox{0.5\textwidth}{
\psfig{figure=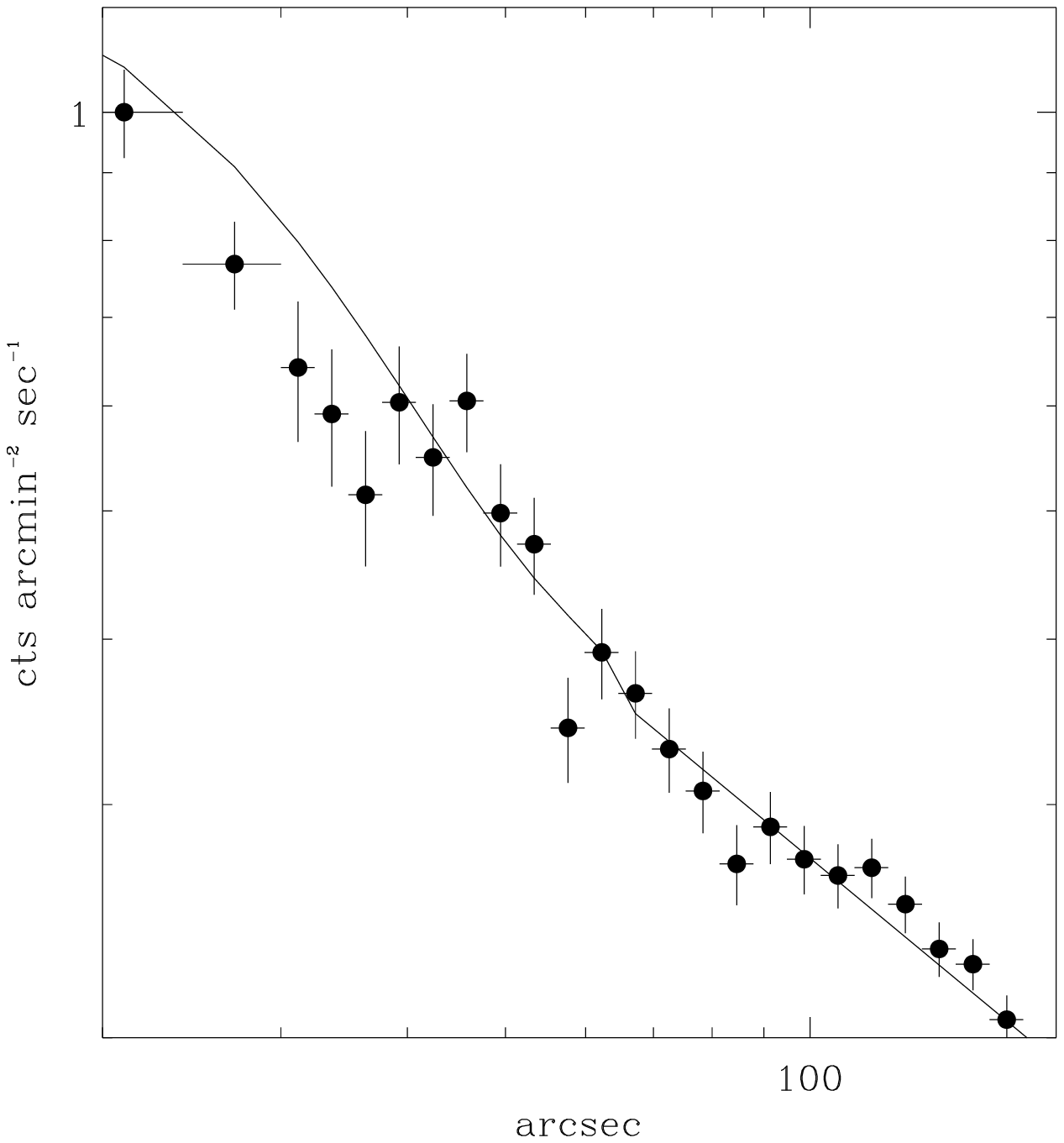,width=0.4\textwidth}}
\parbox{0.5\textwidth}{
\psfig{figure=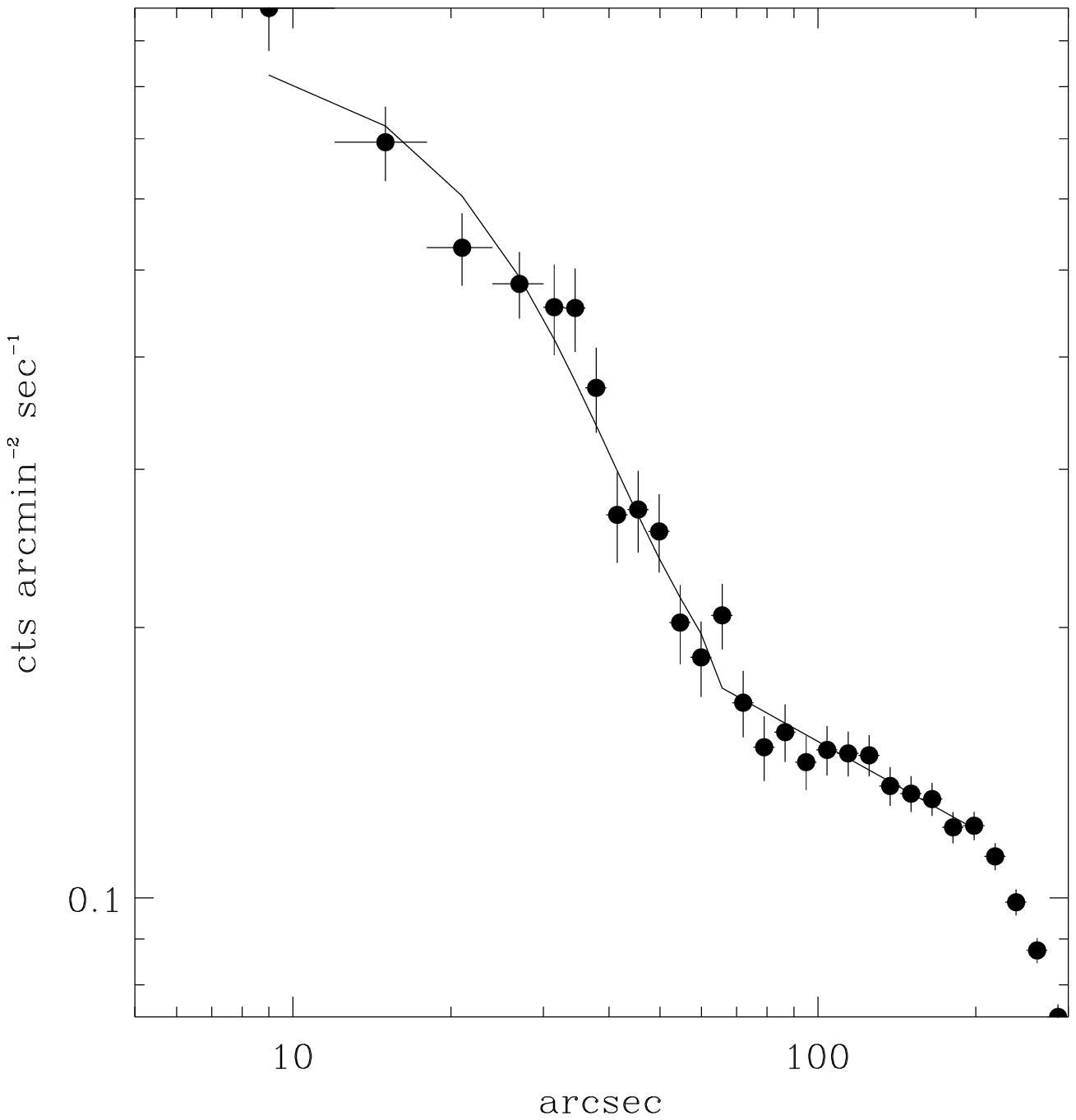,width=0.4\textwidth}}
}
\caption{\label{fig:coldfronts} \footnotesize 
\emph{Left panel:} Summed MOS and pn surface brightness profile across the cold front in the A1 sector of subcluster A
shown in Fig.\ref{fig:sectors}, together with the best-fit model of a power law to describe the outer ICM
and a $\beta$-model to describe the density inside the edge.
\emph{Right panel:} Same as the left panel for the cold front in sector B1 of subcluster B.
}
\end{figure*}

\subsection{The dynamical state of Abell 2028}

The X-ray morphology of the two subclusters associated with Abell 2028, with their
tails and cold fronts, clearly suggests an off-axis merger as seen in various simulations 
\citep[e.g.][]{Ricker.ea:01}.
It is therefore interesting to look for other information about the dynamical state of the cluster
at different wavelengths.
Abell 2028 in the optical band is an elongated structure
as shown by galaxy iso-density images \citep[see Fig. 8 of][]{Trevese.ea:97} or by wavelet analysis 
\citep[see the panels on scales of 750 kpc and 527 kpc of Fig.13 in ][]{Flin.ea:06}. On a scale of
258 kpc, \citet{Flin.ea:06} detected substructures in the 2D galaxy distribution. These structures
are coincident with the X-ray subclusters.
The surface brightness peak of A 2028 A is coincident with the galaxy
GIN 416, the BCG of the subcluster in the optical. The peak of A 2028 B is coincident 
with the pair of galaxies \object{LEDA 140479} and \object{2MASX J15092098+0738225}, which seem to be interacting (see 
the left panel of Fig.\ref{fig.optical}).
The velocities of these three galaxies are $cz=23148$ km s$^{-1}$, $cz=23011$ km s$^{-1}$, and  
$cz=23135$ km s$^{-1}$, suggesting a merger very close to the plane of the sky. 
The definitive confirmation of the presence of a corona has to wait for \chandra\ data with its better PSF; however 
further support is given by GIN 416 also being a radio source, NVSS 150928+073324, 
with $L_{\nu} = 4.1\times10^{23}$ W Hz$^{-1}$ 
(neglecting the small K-correction term), which is consistent with the idea that a radio active BCG should
at least have a corona to be in such a state, because there is no large-scale cool core \citep{Sun:09}.

The galaxy distribution is elongated in the direction of the closest neighbor \object{Abell 2033}. Abell 2028 and
Abell 2033, together with Abell 2029 and \object{Abell 2066}, form the super-cluster number 154 in the catalogue
of \citet{Einasto.ea:01}. It is likely that another cluster can be added to the membership of the supercluster:
the bright \rosat\ source \object{1RXS J150935.9+084605} \citep{Voges.ea:99} is associated with a clustering of 25 SDSS 
galaxies, with measured spectroscopic redshifts, at the redshift of the supercluster (a Gaussian fit to the histogram of 
the galaxies with measured redshift gives $cz=23948$ km s$^{-1}$ corresponding to a redshift of z=0.0799, with a velocity 
dispersion $\sigma=574\pm198$ km s$^{-1}$).
The brightest galaxy in the field, \object{SDSS J150936.33+084632.8}, is a red elliptical whose photometric redshift 
measured by template fitting is $0.0788\pm0.009$. 

The subclusters of A 2028 are therefore merging to produce a more massive cluster belonging to this large-scale 
supercluster. If we coarsely indicate the direction of merging as indicated by the "tails" of the two subclusters
\cite[and by the merging cold fronts perpendicular to them, see][and references therein]{Markevitch.ea:07},
and the direction of the super-cluster as indicated by the four members in the right panel Fig.\ref{fig.optical},
A2028 A and B are merging at $\sim60^{\circ}$ direction with respect to the axis of the super-cluster.
In order to bound in a quantitative way the motion of the two subclusters we determine the jump in gas density at the 
fronts, assuming that the gas density profile is described by the standard $\beta$ model 
$n = n_0 \left[ 1 +\left( r / r_c \right)^2 \right]^{-3/2\beta}$
inside the cold front ($ r > r_{cf}$ where $r_{cf}$ is the cold front radius)
and by a power law $ n = n_1 \left(r\right)^{-\gamma} $ 
outside the cold front ($ r > r_{cf}$). We then calculate the corresponding projected surface brightness.
We fix the cold front radius at $r_{cf} = 65$\arcsec\ for the subclump A and 
$r_{cf} = 65$\arcsec\ for subclump B. We fit the outer part of the surface brightness profile (summing the MOS and pn 
profiles shown in Fig.\ref{fig:sectors}) to set the 
external component parameters, and we successively derive the best fit 
parameters of the innermost part. In Fig.\ref{fig:coldfronts} we plot the surface brightness profiles
with the best-fit models. We find a density discontinuity at the cold front in subcluster A of 
$n_{in}/n_{out} = 3.11 \pm 0.57$,
which combined with the temperature values in the regions inside and outside the front (see Section\ref{sec:1T}), 
leads to a 
pressure ratio $P_{in}/P_{out} = 2.61 \pm 0.53$.
In subclump B, we find a density discontinuity at the cold front 
$n_{in}/n_{out} = 5.86 \pm 1.17$, which combined with the temperature values, leads to a pressure ratio
$P_{in}/P_{out} = 4.31 \pm 0.98$.
Following \citet{Vikhlinin.ea:01}, these pressure ratios correspond to a Mach 
number ${\cal M} = 1.19 ^{+ 0.15}_{-0.18}$ for 
subclump A and ${\cal M} = 1.61^{+0.20}_{-0.22}$ for subclump B respectively.

\section{Conclusions}

We analyzed an \xmm\ observation of Abell 2028, a $\sim4$ keV cluster 
previously known with \asca\ data to display
anomalously high global abundances. The main results of our work can be summarized as follows:

\begin{itemize} 
\item Abell 2028 is actually composed of two subclusters in the process of merging, along the axis of a supercluster.
The evidence suggesting this interpretation is the cometary-like structure of the two subclusters with cold fronts
at the leading edges of the two structures.
The brightest galaxy of the main subcluster is likely hosting a cool corona;
\item The abundance of the clusters, as derived from a temperature and abundance map, is consistent with the
local abundance of more massive clusters. The abundances in every region chosen for spectral analysis show, 
with the appropriate temperature modeling, values in the range 0.3-0.5 \solar, much lower than the 
previous \asca\ global measurement of 0.77 \solar\ \citep[in units of][]{Anders.ea:89};
\item We suggest that the inverse Fe-bias, i.e. the overestimate of iron abundance when fitting 
multi-temperature spectra with average spectroscopic temperature around 3 keV with a single temperature model, is likely 
the source of the high abundances in 3-4 keV local clusters as already discussed in the literature. 
We suggest that the behavior of the Fe-L shell lines explains the bias. The Fe-L complex 
strength falls rapidly with increasing temperature above roughly 3 keV: in a multi-T spectrum where the
average T is near 3 keV, but there are temperature components below and above 3 keV, most of the the Fe
L emission of this multi-T spectrum comes from the lower-T components in the spectrum. The best-fit single temperature 
will be high enough that the Fe L lines will be weaker, so the model will have to compensate by increasing the Fe 
abundance above the true value.

\end{itemize}

More definitive results will be obtained with our ongoing project of observing local clusters in the 3-4 keV range.
More quantitative results for the dynamical state of A 2028 and for the cold fronts and corona detected in the \xmm\ data
will be possible with a dedicated \chandra\ follow-up.

\section*{ACKNOWLEDGEMENTS}
We would like to thank Wayne Baumgartner for providing the \asca\ spectral files and for
useful discussions. We would like to thank the referee, P. Mazzotta, for useful comments and suggestions.
We acknowledge the financial contribution from contracts ASI-INAF
I/023/05/0 and I/088/06/0 and from the NASA grant NNX08AX73G.
We are grateful to the ACE SWEPAM instrument team and the ACE Science center for
providing the ACE data. This research made use of the NASA/IPAC Extragalactic Database (NED) 
which is operated by the Jet Propulsion Laboratory, California Institute of Technology, under contract with the 
National Aeronautics and Space Administration.
Funding for the Sloan Digital Sky Survey (SDSS) has been provided by the Alfred P. Sloan Foundation, 
the Participating Institutions, the National Aeronautics and Space Administration, the National Science Foundation, 
the U.S. Department of Energy, the Japanese Monbukagakusho, and the Max Planck Society. 
The SDSS Web site is http://www.sdss.org/.
The SDSS is managed by the Astrophysical Research Consortium (ARC) for the Participating Institutions. 
The Participating Institutions are The University of Chicago, Fermilab, the Institute for Advanced Study, 
the Japan Participation Group, The Johns Hopkins University, Los Alamos National Laboratory, 
the Max-Planck-Institute for Astronomy (MPIA), the Max-Planck-Institute for Astrophysics (MPA), 
New Mexico State University, University of Pittsburgh, Princeton University, 
the United States Naval Observatory, and the University of Washington.

\bibliographystyle{aa} 
\bibliography{gasta} 

\appendix

\section{Analysis of simulated X-ray spectra}\label{sec.appendix}

In this section we investigate the inverse Fe bias
when fitting simulated multi-temperature models with a single-temperature
plasma model. Following a well-established approach we assumed a double temperature
model (2T) to simulate a plasma with substantial temperature
structure \citep[e.g.][]{Buote:00*1,Balestra.ea:07,Rasia.ea:08}. Even though it may be
argued that a model with a more continuous distribution of emission measure
could be more physically motivated \citep[see for example the discussion for the core of Hydra A
in][]{Simionescu.ea:09}, it is difficult to disctinguish it with respect to a 2T model;
this also seems to be realistic on physical grounds given the possible interpretation of 
a cool corona embedded in an ambient hotter plasma we put forward to explain the core of Abell 2028 A,
which we used as a reference starting point for our simulations.
We perform several simulations of spectra with different assumptions, as
described in detail below, and explore the
possible conditions that can potentially affect the distribution of
best-fit values of $kT$ and, most important, of the metallicity $Z$.

We performed a set of simulations of clusters with two temperature components (specified
in Table \ref{tab:sim}) in the range relevant for the core of Abell 2028 A, i.e. 1.3-2 keV
for the cold component and 4-6 keV for the hot component with an emission measure ratio
(the normalization of the APEC model) ranging from 0.1 to 1.0 with steps of 0.1 between the two
temperature components. 
The starting dominant component is listed as first, with an initial normalization (in XSPEC units) 
of $1.5\times10^{-4}$. The total number of counts change as a function of the ratio 
$\rm{EM_{cold}/EM_{hot}}$ and ranges from $\sim$1100 counts and $\sim$2600 for MOS and pn in the 0.5-10 keV band 
in the simulations with $\rm{EM_{cold}/EM_{hot}}=0.1$ to $\sim$1800 and $\sim$4600 for MOS and pn
in the simulations with $\rm{EM_{cold}/EM_{hot}}=1.0$.
For each set of simulations, we performed 20 realizations and listed the results of the fits with a 1T model
in the various table of results as median and mean absolute deviation from the median. We simulated spectra 
with the Galactic column density and redshift of A 2028 with an exposure time of 50 ks.
We also investigated the distribution of best-fit abundances in the presence of a gradient in the
abundance profile, with a higher abundance associated with the lower temperature component.
We quote and plot the results separately for the three EPIC detectors to investigate the presence of 
effects arising by the difference in effective area. 

\begin{figure}
\centerline{
\hspace{6.5truecm}
 \epsfig{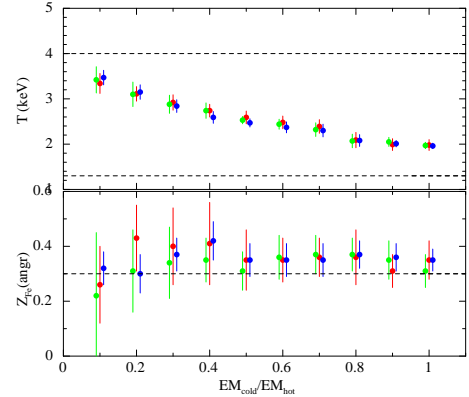} }
\caption{ Median of the best-fit temperature and abundance distribution as a function of the ratio of the cold and hot 
emission measure, for 
the simulation s1. The green dots refer to M1 results, the red dots to M2 and the blue dots to pn. The dashed lines 
refer to the input values of temperature and abundance for the two components. The error bars correspond to the mean 
absolute deviation.
} \label{fig:sim_s1}
\end{figure}

\begin{figure}
\centerline{
\hspace{6.5truecm}
 \epsfig{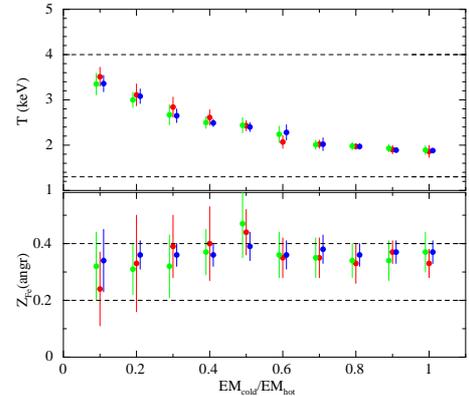} }
\caption{Same as in Fig.\ref{fig:sim_s1}  for the simulations s2.
Note the two lines in the abundance panel which refer to the two different input abundances in the
two components: the higher one (0.4 \solar) in the cold component and the lower
one (0.2 \solar) in the hot component.  
} \label{fig:sim_s3}
\end{figure}

\begin{figure}
\centerline{
\hspace{6.5truecm}
 \epsfig{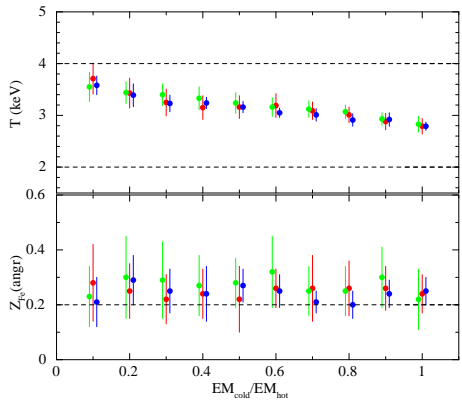} }
\caption{Same as in Fig.\ref{fig:sim_s1}  for the simulations s3.
 } \label{fig:sim_s5}
\end{figure}

\begin{figure}
\centerline{
\hspace{6.5truecm}
 \epsfig{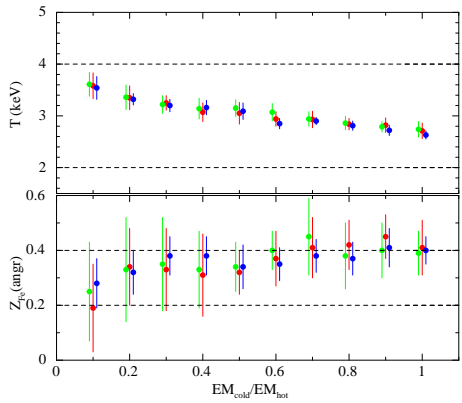} }
\caption{Same as in Fig.\ref{fig:sim_s1}  for the simulations s4.
 } \label{fig:sim_s6}
\end{figure}

\begin{figure}
\centerline{
\hspace{6.5truecm}
 \epsfig{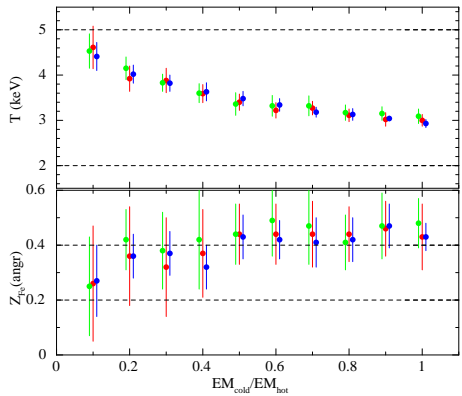} }
\caption{Same as in Fig.\ref{fig:sim_s1}  for the simulations s5.
 } \label{fig:sim_s7}
\end{figure}

\begin{figure}
\centerline{
\hspace{6.5truecm}
 \epsfig{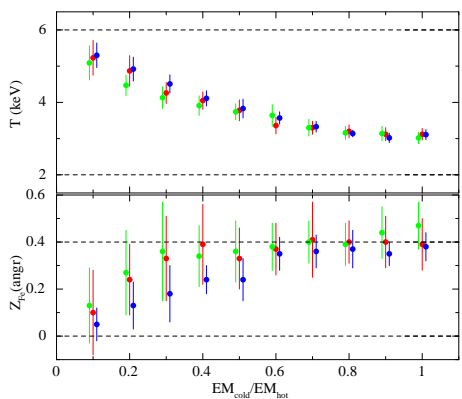} }
\caption{Same as in Fig.\ref{fig:sim_s1}  for the simulations s6.
 } \label{fig:sim_s8}
\end{figure}

\begin{figure}
\centerline{
 \epsfig{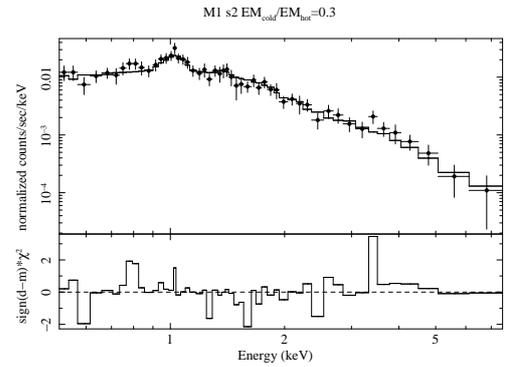} }
\caption{The MOS 1 spectrum and best fit 1T model with residuals for a realization
of the simulation s1 with $\rm{EM_{cold}/EM_{hot}}=0.3$. The input model
consists of two plasmas with $\rm{kT_{cold}}=1.3$ keV and $\rm{kT_{hot}}=4$ keV with an abundance 
of 0.3 \solar\ in both components, the single-temperature model
best fit values are $\rm{kT}=3.02\pm0.22$ keV and  $\rm{Z}=0.72\pm0.23$ \solar.
 } \label{fig:sim_febias}
\end{figure}

\begin{figure}
\centerline{
 \epsfig{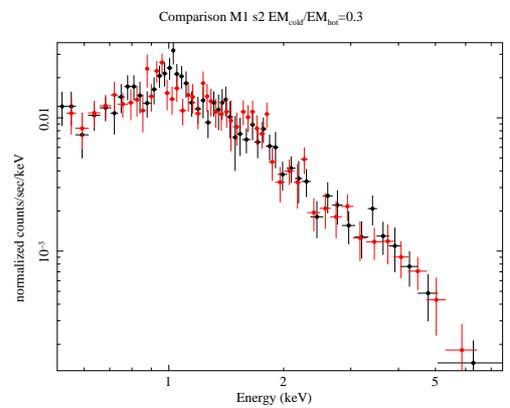} }
\caption{Comparison of the MOS 1 spectra for two different realizations
of the simulation s1 with $\rm{EM_{cold}/EM_{hot}}=0.3$: the black dots refer to the 
realization of Fig.\ref{fig:sim_febias} and the red dots to a realization with best
fit abundance of 0.07 \solar.
 } \label{fig:sim_febias_comparison}
\end{figure}

It can be seen by the plots of the results for the various simulations (from Fig. \ref{fig:sim_s1}
to \ref{fig:sim_s8}), which are also listed in Table \ref{tab:sim_res}, that, as the contribution
of the low temperature component increases, the best-fit temperature of the single component model
shifts to lower values. The distribution of the best fit abundance, either in cases with the same input
abundance for the two components or in cases with an higher abundance in the lower temperature component,
is always biased to higher values that a naive emission measure weighting would predict, up to $\sim50$\% of the
median value compared to the input value in some cases. Even more striking is the large scatter in the distributions in 
particular for low values of the ratio $\rm{EM_{cold}/EM_{hot}}$ and for the MOS detectors, again pointing 
to the relevance of the Fe-L line in this bias: the fitting of the feature is key for the abundance determination,
and Poisson noise is limiting for lower effective area instruments. Rather extreme values may occur: we show from 
example a MOS 1 s1 simulation when  $\rm{EM_{cold}/EM_{hot}}=0.3$ (see Fig.\ref{fig:sim_febias}): the best-fit 
abundance is $0.72\pm0.23$ \solar\ with a statistically good fit ($\chi^2/\rm{dof}<1$) with just a single temperature 
(we recall that the input abundance value for both components is 0.3 \solar). It is instructive to
plot this example together with another realization of the same model, with a best-fit abundance of 
$0.07^{+0.12}_{-0.07}$ \solar, because it shows the importance of the fitting of the spectral region of the Fe-L 
lines (see Fig.\ref{fig:sim_febias_comparison}). In both cases the best fit temperatures are around $\sim3$ keV.
That the role played by the contribution of the Fe-K line is not relevant in the inverse Fe bias is also
suggested by the simulation s6, where we only use a bremsstrahlung (i.e. the abundance was set to zero) for the
hot component. Already for $\rm{EM_{cold}/EM_{hot}}=0.4$ the median best fit abundance is equal to the input abundance 
of the cold component.

\begin{figure}
\centerline{
\hspace{6.5truecm}
 \epsfig{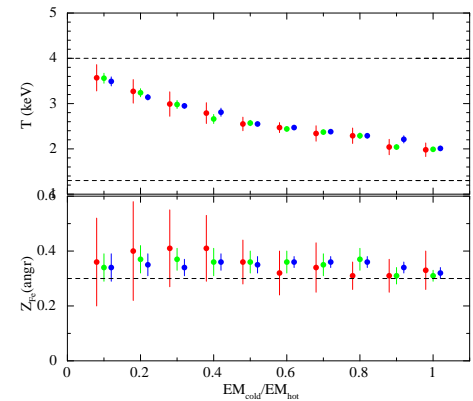} }
\caption{Median of the best-fit temperature and abundance distribution as a function of the ratio of the cold and hot 
emission measure, for the simulation s1 for the detector MOS2, with the same number counts for all the $\rm{EM_{cold}/EM_{hot}}$ ratios.
The red dots refer to spectra with $1\times10^{3}$ counts in the 0.5-10.0 keV band, the green ones to spectra with
$5\times10^{3}$ counts, and the blue to spectra with $1\times10^{4}$ counts.
The dashed lines refer to the input values of temperature and abundance for the two components. The error bars correspond 
to the mean absolute deviation.
 } \label{fig:sim_counts}
\end{figure}

The trends shown here do not change if one instead selects spectra with the same number counts as the ratio 
$\rm{EM_{cold}/EM_{hot}}$ increases. As an example we show the results in Fig.\ref{fig:sim_counts} of the simulation s1 for 
MOS2 for three different number counts for all the $\rm{EM_{cold}/EM_{hot}}$ ratios, i.e. $1\times10^{3}$, $5\times10^{3}$ 
and $1\times10^{4}$ counts in the 0.5-10 keV band.

Another result coming from the investigation of the simulations is that restricting the fitting
energy band to just fit the shape of the Fe-L bump (for example in the 0.5-3 keV energy band) is no guarantee
of recovering the abundance of the low-temperature component: the fitted continuum is too high because of the
contribution of the high-temperature component and the abundance thus fitted is too low (see the green and blue
pn spectra of Fig.\ref{fig:speccorona} for a visual representation).

\begin{table}
\caption{Input values for the spectral simulations of two temperature
components as described in the Appendix}
\label{tab:sim}
\centering
\begin{tabular}{ l l l }
\hline\hline
Bim${^a}$ & $kT_{inp}\mathrm{^b}$ & $Z_{inp}\mathrm{^c}$ \\
\hline
s1 &  4+1.3 & 0.3 \\
s2 &  4+1.3 & 0.2+0.4 \\
s3 &  4+2 & 0.2 \\
s4 &  4+2 & 0.2+0.4 \\
s5 &  5+2 & 0.2+0.4 \\
s6 &  6+2 & 0.0+0.4 \\
\hline
\end{tabular}
\begin{list}{}{}
\item[Notes:] $\mathrm{^a}$ simulations identification number;
$\mathrm{^b}$ input temperature of the hot and cold component;
$\mathrm{^c}$ input iron abundance;
$\mathrm{^c}$ XSPEC fixed norm of the first temperature component ($\rm{cm}^{-5}$);
\end{list}
\end{table}

%
\clearpage
\newpage
\begin{table*}
\scriptsize
\caption{Fitted values with a 1T model in the 0.5-10 keV band for the three EPIC detectors of the simulations listed in
Table \ref{tab:sim}.}
\label{tab:sim_res}
\centering
\begin{tabular}{c c c c c c c c c c c}
\hline\hline
& Ratio 0.1 & 0.2 & 0.3 & 0.4 & 0.5 & 0.6 & 0.7 & 0.8 & 0.9 & 1.0\\
\hline  
&           &     &     &      & s1 & M1  &     &     &     &     \\ 
\hline
kT & $3.34\pm0.22$ & $3.11\pm0.16$ & $2.92\pm0.17$ & $2.74\pm0.14$ & $2.59\pm0.14$ & $2.48\pm0.14$ & $2.39\pm0.15$ & $2.09\pm0.17$ & $1.99\pm0.13$ & $1.98\pm0.12$ \\ 
Z &  $0.26\pm0.14$ & $0.43\pm0.12$ & $0.40\pm0.14$ & $0.41\pm0.15$ & $0.35\pm0.11$ & $0.35\pm0.08$ & $0.36\pm0.07$ & $0.36\pm0.10$ & $0.31\pm0.06$ & $0.35\pm0.07$ \\
\hline
&           &     &     &      & s1 & M2  &     &     &     &     \\ 
\hline
kT & $3.42\pm0.29$ & $3.10\pm0.27$ & $2.88\pm0.20$ & $2.74\pm0.17$ & $2.53\pm0.08$ & $2.44\pm0.11$ & $2.32\pm0.15$ & $2.07\pm0.15$ & $2.05\pm0.10$ & $1.97\pm0.07$ \\
Z &  $0.22\pm0.23$ & $0.31\pm0.15$ & $0.34\pm0.13$ & $0.35\pm0.08$ & $0.31\pm0.07$ & $0.36\pm0.08$ & $0.37\pm0.07$ & $0.37\pm0.06$ & $0.35\pm0.07$ & $0.31\pm0.06$ \\
\hline
&           &     &     &      & s1 & pn  &     &     &     &     \\ 
\hline
kT & $3.47\pm0.16$ & $3.15\pm0.16$ & $2.84\pm0.14$ & $2.59\pm0.13$ & $2.47\pm0.09$ & $2.37\pm0.12$ & $2.30\pm0.14$ & $2.08\pm0.13$ & $2.01\pm0.07$ & $1.96\pm0.06$ \\
Z &  $0.32\pm0.06$ & $0.30\pm0.07$ & $0.37\pm0.06$ & $0.42\pm0.07$ & $0.35\pm0.06$ & $0.35\pm0.06$ & $0.35\pm0.06$ & $0.37\pm0.05$ & $0.36\pm0.05$ & $0.35\pm0.04$ \\
\hline
&           &     &     &      & s2 & M1  &     &     &     &     \\ 
\hline
kT & $3.51\pm0.21$ & $3.11\pm0.24$ & $2.84\pm0.22$ & $2.61\pm0.17$ & $2.43\pm0.11$ & $2.07\pm0.14$ & $2.02\pm0.09$ & $1.97\pm0.06$ & $1.90\pm0.09$ & $1.86\pm0.13$ \\
Z &  $0.24\pm0.13$ & $0.33\pm0.17$ & $0.39\pm0.11$ & $0.40\pm0.13$ & $0.44\pm0.08$ & $0.35\pm0.07$ & $0.35\pm0.07$ & $0.33\pm0.07$ & $0.37\pm0.04$ & $0.33\pm0.05$ \\
\hline
&           &     &     &      & s2 & M2  &     &     &     &     \\ 
\hline
kT & $3.35\pm0.24$ & $3.00\pm0.17$ & $2.67\pm0.22$ & $2.50\pm0.13$ & $2.44\pm0.17$ & $2.24\pm0.18$ & $2.01\pm0.10$ & $1.98\pm0.08$ & $1.93\pm0.08$ & $1.89\pm0.10$ \\
Z &  $0.32\pm0.12$ & $0.31\pm0.09$ & $0.32\pm0.11$ & $0.37\pm0.08$ & $0.47\pm0.12$ & $0.36\pm0.08$ & $0.35\pm0.07$ & $0.34\pm0.06$ & $0.34\pm0.07$ & $0.37\pm0.07$ \\
\hline
&           &     &     &      & s2 & pn  &     &     &     &     \\ 
\hline
kT & $3.36\pm0.18$ & $3.08\pm0.16$ & $2.65\pm0.15$ & $2.49\pm0.08$ & $2.40\pm0.10$ & $2.28\pm0.17$ & $2.02\pm0.14$ & $1.97\pm0.06$ & $1.89\pm0.03$ & $1.88\pm0.03$ \\
Z & $0.34\pm0.11$ & $0.36\pm0.05$ & $0.36\pm0.04$ & $0.36\pm0.04$ & $0.39\pm0.05$ & $0.36\pm0.05$ & $0.38\pm0.05$ & $0.36\pm0.04$ & $0.37\pm0.04$ & $0.37\pm0.04$ \\
\hline
&           &     &     &      & s3 & M1  &     &     &     &     \\ 
\hline
kT & $3.71\pm0.30$ & $3.43\pm0.29$ & $3.25\pm0.26$ & $3.15\pm0.23$ & $3.16\pm0.22$ & $3.19\pm0.23$ & $3.09\pm0.17$ & $3.01\pm0.15$ & $2.88\pm0.16$ & $2.79\pm0.15$ \\
Z &  $0.28\pm0.14$ & $0.25\pm0.10$ & $0.22\pm0.09$ & $0.24\pm0.09$ & $0.22\pm0.12$ & $0.26\pm0.07$ & $0.26\pm0.12$ & $0.26\pm0.10$ & $0.26\pm0.08$ & $0.24\pm0.07$ \\
\hline
&           &     &     &      & s3 & M2  &     &     &     &     \\ 
\hline
kT & $3.55\pm0.28$ & $3.44\pm0.21$ & $3.40\pm0.21$ & $3.33\pm0.22$ & $3.24\pm0.20$ & $3.16\pm0.18$ & $3.12\pm0.16$ & $3.07\pm0.13$ & $2.93\pm0.12$ & $2.83\pm0.15$ \\
Z &  $0.23\pm0.11$ & $0.30\pm0.15$ & $0.29\pm0.14$ & $0.27\pm0.11$ & $0.28\pm0.09$ & $0.32\pm0.13$ & $0.25\pm0.09$ & $0.25\pm0.09$ & $0.30\pm0.11$ & $0.22\pm0.11$ \\
\hline
&           &     &     &      & s3 & pn  &     &     &     &     \\ 
\hline
kT & $3.58\pm0.18$ & $3.39\pm0.22$ & $3.23\pm0.16$ & $3.24\pm0.11$ & $3.16\pm0.11$ & $3.05\pm0.09$ & $3.01\pm0.12$ & $2.91\pm0.12$ & $2.92\pm0.13$ & $2.79\pm0.08$ \\
Z &  $0.21\pm0.09$ & $0.29\pm0.09$ & $0.25\pm0.08$ & $0.24\pm0.10$ & $0.27\pm0.06$ & $0.25\pm0.06$ & $0.21\pm0.04$ & $0.20\pm0.05$ & $0.24\pm0.05$ & $0.25\pm0.05$ \\
\hline
&           &     &     &      & s4 & M1  &     &     &     &     \\ 
\hline
kT & $3.58\pm0.25$ & $3.35\pm0.23$ & $3.25\pm0.14$ & $3.07\pm0.18$ & $3.05\pm0.21$ & $2.94\pm0.14$ & $2.93\pm0.16$ & $2.84\pm0.11$ & $2.82\pm0.14$ & $2.71\pm0.15$ \\
Z &  $0.19\pm0.16$ & $0.34\pm0.14$ & $0.33\pm0.15$ & $0.31\pm0.15$ & $0.32\pm0.08$ & $0.37\pm0.10$ & $0.41\pm0.11$ & $0.42\pm0.09$ & $0.45\pm0.08$ & $0.41\pm0.10$ \\
\hline
&           &     &     &      & s4 & M2  &     &     &     &     \\ 
\hline
kT & $3.61\pm0.23$ & $3.36\pm0.24$ & $3.22\pm0.17$ & $3.14\pm0.20$ & $3.15\pm0.16$ & $3.07\pm0.17$ & $2.94\pm0.14$ & $2.86\pm0.13$ & $2.79\pm0.10$ & $2.74\pm0.15$ \\
Z &  $0.25\pm0.18$ & $0.33\pm0.19$ & $0.35\pm0.17$ & $0.33\pm0.14$ & $0.34\pm0.09$ & $0.40\pm0.07$ & $0.45\pm0.14$ & $0.38\pm0.12$ & $0.40\pm0.10$ & $0.39\pm0.08$ \\
\hline
&           &     &     &      & s4 & pn  &     &     &     &     \\ 
\hline
kT & $3.54\pm0.22$ & $3.32\pm0.11$ & $3.20\pm0.12$ & $3.16\pm0.14$ & $3.09\pm0.16$ & $2.85\pm0.10$ & $2.90\pm0.07$ & $2.81\pm0.09$ & $2.72\pm0.10$ & $2.63\pm0.08$ \\
Z &  $0.28\pm0.09$ & $0.32\pm0.08$ & $0.38\pm0.07$ & $0.38\pm0.07$ & $0.34\pm0.08$ & $0.35\pm0.06$ & $0.38\pm0.06$ & $0.37\pm0.06$ & $0.41\pm0.07$ & $0.40\pm0.05$ \\
\hline
&           &     &     &      & s5 & M1  &     &     &     &     \\ 
\hline
kT & $4.61\pm0.47$ & $3.92\pm0.28$ & $3.88\pm0.27$ & $3.59\pm0.20$ & $3.40\pm0.18$ & $3.22\pm0.17$ & $3.27\pm0.15$ & $3.11\pm0.14$ & $3.02\pm0.15$ & $3.00\pm0.13$ \\
Z &  $0.26\pm0.21$ & $0.36\pm0.18$ & $0.32\pm0.18$ & $0.37\pm0.16$ & $0.44\pm0.11$ & $0.44\pm0.11$ & $0.44\pm0.12$ & $0.44\pm0.10$ & $0.46\pm0.10$ & $0.43\pm0.12$ \\
\hline
&           &     &     &      & s5 & M2  &     &     &     &     \\ 
\hline
kT & $4.53\pm0.38$ & $4.15\pm0.25$ & $3.83\pm0.19$ & $3.60\pm0.21$ & $3.36\pm0.25$ & $3.32\pm0.23$ & $3.32\pm0.22$ & $3.17\pm0.17$ & $3.15\pm0.15$ & $3.09\pm0.16$ \\
Z &  $0.25\pm0.18$ & $0.42\pm0.11$ & $0.38\pm0.14$ & $0.42\pm0.18$ & $0.44\pm0.11$ & $0.49\pm0.13$ & $0.47\pm0.14$ & $0.41\pm0.10$ & $0.47\pm0.12$ & $0.48\pm0.09$ \\
\hline
&           &     &     &      & s5 & pn  &     &     &     &     \\ 
\hline
kT & $4.41\pm0.31$ & $4.02\pm0.20$ & $3.82\pm0.18$ & $3.63\pm0.20$ & $3.48\pm0.16$ & $3.34\pm0.14$ & $3.18\pm0.11$ & $3.13\pm0.13$ & $3.04\pm0.05$ & $2.93\pm0.09$ \\
Z &  $0.27\pm0.13$ & $0.36\pm0.08$ & $0.37\pm0.08$ & $0.32\pm0.08$ & $0.43\pm0.08$ & $0.42\pm0.07$ & $0.41\pm0.09$ & $0.42\pm0.08$ & $0.47\pm0.08$ & $0.43\pm0.05$ \\
\hline
&           &     &     &      & s6 & M1  &     &     &     &     \\ 
\hline
kT & $5.23\pm0.48$ & $4.87\pm0.42$ & $4.26\pm0.29$ & $4.05\pm0.24$ & $3.78\pm0.29$ & $3.36\pm0.23$ & $3.30\pm0.18$ & $3.20\pm0.18$ & $3.12\pm0.18$ & $3.12\pm0.16$ \\
Z &  $0.10\pm0.18$ & $0.24\pm0.15$ & $0.33\pm0.18$ & $0.39\pm0.17$ & $0.33\pm0.13$ & $0.37\pm0.11$ & $0.41\pm0.16$ & $0.40\pm0.09$ & $0.40\pm0.11$ & $0.39\pm0.11$ \\
\hline
&           &     &     &      & s6 & M2  &     &     &     &     \\ 
\hline
kT & $5.09\pm0.47$ & $4.47\pm0.28$ & $4.13\pm0.30$ & $3.91\pm0.27$ & $3.74\pm0.22$ & $3.64\pm0.30$ & $3.30\pm0.23$ & $3.16\pm0.18$ & $3.14\pm0.20$ & $3.02\pm0.16$ \\
Z &  $0.13\pm0.16$ & $0.27\pm0.18$ & $0.36\pm0.21$ & $0.34\pm0.13$ & $0.36\pm0.13$ & $0.38\pm0.10$ & $0.40\pm0.09$ & $0.39\pm0.09$ & $0.44\pm0.11$ & $0.47\pm0.10$ \\
\hline
&           &     &     &      & s6 & pn  &     &     &     &     \\ 
\hline
kT & $5.30\pm0.34$ & $4.92\pm0.33$ & $4.51\pm0.25$ & $4.11\pm0.21$ & $3.83\pm0.26$ & $3.57\pm0.17$ & $3.33\pm0.15$ & $3.14\pm0.09$ & $3.02\pm0.13$ & $3.11\pm0.14$ \\
Z &  $0.05\pm0.07$ & $0.13\pm0.10$ & $0.18\pm0.12$ & $0.24\pm0.06$ & $0.24\pm0.09$ & $0.35\pm0.07$ & $0.36\pm0.07$ & $0.37\pm0.08$ & $0.35\pm0.05$ & $0.38\pm0.06$ \\
\hline
\end{tabular}
\end{table*}
\clearpage

\end{document}